\documentclass{elsartL}
\usepackage{graphicx}
\usepackage{amssymb}
\usepackage{amsmath}
\usepackage{physics}
\usepackage{mathrsfs}
\usepackage{mathtools}
\usepackage{appendix}
\newcommand{\avg}[1]{\left< #1 \right>} % For a nice display of the average value
\begin{document}

\begin{frontmatter}
\title{Determination of the masses and decay widths of the scalar-isoscalar and vector-isovector mesons below $2$ GeV}
\author{Evangelos Matsinos}

\begin{abstract}This study revisits the issue of the extraction of estimates for the masses and for the partial decay widths to two pions of the scalar-isoscalar $I^G\,(J^{PC}) = 0^+\,(0^{++})$ and vector-isovector 
$I^G\,(J^{PC}) = 1^+\,(1^{--})$ mesons below $2$ GeV, from the data contained in the recent compilation by the Particle Data Group.\\
\noindent {\it PACS 2010:} 13.20.Jf, 13.25.-k, 13.25.Jx, 14.40.-n, 14.40.Be
%
%13.20.Jf Decays of other mesons
%13.25.-k Hadronic decays of mesons
%13.25.Jx Decays of other mesons
%14.40.-n Properties of mesons
%14.40.Be Properties of light mesons
%
\end{abstract}
\end{frontmatter}

\section{\label{sec:Introduction}Introduction}

The main motivation of this work stems from the application of its results in a pion-nucleon ($\pi N$) interaction model which is based on $s$-, $u$-, and $t$-channel Feynman graphs \cite{matsinos2017}. In the context of this 
model, the $t$-channel contributions to the strong-interaction part of the $s$- and $p$-wave scattering amplitudes originate from exchanges of scalar-isoscalar $I^G\,(J^{PC}) = 0^+\,(0^{++})$ and vector-isovector 
$I^G\,(J^{PC}) = 1^+\,(1^{--})$ mesons with rest masses below $2$ GeV. Until now, the recommendations by the Particle Data Group (PDG) for the masses and for the partial decay widths (to two pions) of these unstable particles 
(resonances) were followed. To include in the model one scalar-isoscalar state, the $f_0(1710)$, required the extraction of estimates for the physical properties of that resonance; no relevant results are given in the recent 
PDG compilation \cite{pdg2020}. As the PDG reports also contain the lists of the data they use (or refrain from using) in their analyses, it was decided to revisit the determination of the physical properties of all 
scalar-isoscalar and vector-isovector resonances (which are relevant in the context of Ref.~\cite{matsinos2017}), rather than unreservedly rely on the PDG recommendations.

The structure of this paper is as follows. The subsequent section provides the details on four statistical methods for extracting estimates from the available data. For didactical reasons, the method using the standard $\chi^2$ 
minimisation function is described first, followed by the introduction of some options towards robust optimisation. Two of the robust methods detailed in Section \ref{sec:MethodsII}, the most insensitive to the presence of 
outliers, will yield the main results (and the recommended values) of this study. Section \ref{sec:Methods} concludes with the description of a fourth analysis method, the one featuring the determination of the properties 
of the resonances from the Cumulative Distribution Function of the relevant physical quantity, obtained directly from the input data. Containing the essential details about the data analysis, Section \ref{sec:Results} is 
divided into two parts: the first part appertains to the scalar-isoscalar resonances, whereas the second deals with the vector-isovector ones. The conclusions of this paper are given in its last section, Section \ref{sec:Conclusions}.

In this work, all masses will be expressed in energy units, namely in MeV; DoF will stand for `degree(s) of freedom', whereas NDF will denote `the number of degrees of freedom'; PDF will stand for `Probability Density Function' 
and CDF for `Cumulative Distribution Function'; MF will be an abbreviation of `minimisation function'.

\section{\label{sec:Methods}Methods}

\subsection{\label{sec:MethodsI}Standard $\chi^2$ MF}

The routine approach to extracting a meaningful average and a relevant uncertainty from a set of $N$ input values $y_i$ with uncertainties $\delta y_i \neq 0$, corresponding to independent observations of one measurable 
quantity, is to introduce a function $\chi^2$ as
\begin{equation} \label{eq:EQ01}
\chi^2 = \sum_{i=1}^{N} \left( \frac{y_i - \avg{y}}{\delta y_i} \right)^2 \equiv \sum_{i=1}^{N} w_i (y_i - \avg{y})^2 \, \, \, ,
\end{equation}
where the weights $w_i \coloneqq (\delta y_i)^{-2}$, and minimise it with respect to the free parameter $\avg{y}$. If the input uncertainties are symmetrical, the solution is obtained analytically:
\begin{equation} \label{eq:EQ02}
\frac{d \chi^2}{d\avg{y}} = -2 \sum_{i=1}^{N} w_i (y_i - \avg{y}) = 0 \Rightarrow \avg{y} = \frac{\sum_{i=1}^{N} w_i y_i}{\sum_{i=1}^{N} w_i} \, \, \, ,
\end{equation}
whereas a numerical solution is sought in case of asymmetrical input uncertainties, via the application of standard minimisation packages like, for the sake of one notable example, the MINUIT software package of the CERN 
library \cite{James1998,James2004}. (MINUIT, which is available in two versions - FORTRAN and C/C++, will be used in the numerical optimisation.) For symmetrical input uncertainties, one obtains the minimal value of the 
$\chi^2$ function as
\begin{equation} \label{eq:EQ03}
\chi^2_{\rm min} = \sum_{i=1}^{N} w_i y_i^2 - \frac{\left( \sum_{i=1}^{N} w_i y_i \right)^2}{\sum_{i=1}^{N} w_i} \, \, \, .
\end{equation}

The fitted uncertainty of $\avg{y}$ is defined as the change $\delta \avg{y}$ (in $\avg{y}$) which increases the $\chi^2_{\rm min}$ value of Eq.~(\ref{eq:EQ03}) by $1$: $\chi^2_{\rm min} \to \chi^2_{\rm min}+1$. The 
$\delta \avg{y}$ value, achieving this result, reads as
\begin{equation} \label{eq:EQ04}
\delta \avg{y} = \frac{1}{\sqrt{\sum_{i=1}^{N} w_i}} \, \, \, .
\end{equation}
The fitted uncertainty, obtained with the MINUIT method MINOS, corresponds to $\delta \avg{y}$ of Eq.~(\ref{eq:EQ04}).

The quality of the description of the input dataset by the constant $\avg{y}$ is judged on the basis of the $\chi^2_{\rm min}$ value of Eq.~(\ref{eq:EQ03}) and the NDF in the problem, namely the number of input datapoints 
$N$ reduced by $1$, as one free parameter is used in Eq.~(\ref{eq:EQ01}). Formally, the assessment of the quality of each fit rests upon the evaluation of the so-called p-value
\begin{equation} \label{eq:EQ05}
{\rm p} (\chi^2_{\rm min}, {\rm NDF}) \coloneqq \int_{\chi^2_{\rm min}}^\infty f(u; {\rm NDF}) du \, \, \, ,
\end{equation}
where $f(u; \nu)$ is the PDF of the $\chi^2$ distribution
\begin{equation} \label{eq:EQ06}
f(u; \nu) = \frac{u^{\nu/2-1} e^{-u/2}}{2^{\nu/2} \Gamma(\nu/2)}
\end{equation}
and $\Gamma(x)$ denotes the standard gamma function. Being the upper tail of the CDF of the $\chi^2$ distribution, the p-value reflects the rarity of the $\chi^2_{\rm min}$ result: if ${\rm p} (\chi^2_{\rm min}, {\rm NDF})$ 
comes out sufficiently `small' (this matter will be addressed shortly), it is unlikely that the re-investigation of the phenomenon (resulting in new input values and uncertainties) will yield $(\chi^2_{\rm min})_{\rm new} > \chi^2_{\rm min}$ 
on the basis of pure chance (i.e., due to statistical fluctuations), hence the $\chi^2_{\rm min}$ result is considered to be `large'.

To accept or reject the null hypothesis (i.e., that the input dataset can be described by one meaningful constant value in the case treated in this work), one first needs to define what `statistical significance' signifies. 
In practice, one introduces a minimal p-value, ${\rm p}_{\rm min}$, for the acceptance of the null hypothesis; ${\rm p}_{\rm min}$, a subjective threshold, is known as `significance level' and is usually denoted in the 
literature by $\alpha$. If the resulting p-value of Eq.~(\ref{eq:EQ05}) comes out (in a fit) short of ${\rm p}_{\rm min}$, then the null hypothesis (in that fit) may be rejected. The ${\rm p}_{\rm min}$ value, which is 
currently accepted by most statisticians as representing the outset of statistical significance, is $1.00 \cdot 10^{-2}$; this threshold will also be adopted in this work. (In some analyses, the threshold $5.00 \cdot 10^{-2}$ 
is taken to indicate \emph{probable} significance, suggesting further investigation.)

Before proceeding, I will define one quantity which will be useful in the discussion: the Birge factor of a fit, introduced by Birge (1887-1980) in 1932 \cite{Birge1932} and mundanely called `scale factor' by the PDG, is 
defined as
\begin{equation} \label{eq:EQ07}
s = \sqrt{\frac{\chi^2_{\rm min}}{\rm NDF}} \, \, \, .
\end{equation}
(The scale factors, reported in the PDG compilation, are not always the Birge factors of this work, i.e., values obtained via Eq.~(\ref{eq:EQ07}). In the Section `Procedures' in the Chapter `Introduction' of the PDG 
compilation one reads: ``When combining data with widely varying errors, we modify this procedure [i.e., the determination of the scale factor via Eq.~(\ref{eq:EQ07})] slightly. We evaluate $S$ [i.e., their scale factor] 
using only the experiments with smaller errors.'')

Despite the fact that the proper measure of the fit quality is the p-value of the fit, the majority of Physics studies favour the use of the reduced $\chi^2_{\rm min}$ value, $\chi^2_{\rm min}/{\rm NDF}$, as such a measure: 
as long as $\chi^2_{\rm min}/{\rm NDF}$ remains close to $1$, the fit is claimed to be satisfactory; of course, it is unclear what `close to' implies. Evidently, the fit is \emph{un}satisfactory when $\chi^2_{\rm min}/{\rm NDF} \not\approx 1$, 
but this condition conveys no information because neither can it tell one `how much' unsatisfactory a fit is on the basis of a given $\chi^2_{\rm min}/{\rm NDF}$ result, nor does it specify the domain of the $\chi^2_{\rm min}/{\rm NDF}$ 
values which are covered by the symbol `$\not\approx$'. (Importantly, the condition $\chi^2_{\rm min}/{\rm NDF} \approx 1$ as a criterion for the acceptance of the null hypothesis is plain \emph{wrong} for small NDF. For 
instance, the result $\chi^2_{\rm min}/{\rm NDF} = 3$ for $5$ DoF is acceptable by most statisticians as ${\rm p} (15, 5) > {\rm p}_{\rm min} = 1.00 \cdot 10^{-2}$.)

The condition $\chi^2_{\rm min}/{\rm NDF} \approx 1$, equivalently $s \approx 1$, is a coarse `rule of thumb' for large NDF. To make reliable use of the $\chi^2_{\rm min}/{\rm NDF}$ value in hypothesis testing, one would have 
to generate curves of the critical $\chi^2_{\rm min}/{\rm NDF}$ versus the NDF for all ${\rm p}_{\rm min}$ values used in an analysis. However, common sense suggests that such an introduction of complexity is pointless. For 
the purposes of hypothesis testing, the extraction of the p-value via Eq.~(\ref{eq:EQ05}) (and its subsequent comparison with the accepted ${\rm p}_{\rm min}$ value) is straightforward, as is easy to implement.

Let me now examine what needs to be done so that the result $\avg{y} \pm \delta \avg{y}$ be representative of the entire input dataset ($y_i,\delta y_i : i = 1, 2, \dots , N$), even when the $\chi^2_{\rm min}/{\rm NDF}$ value 
exceeds the statistical expectation of $1$. Provided that the input data ($y_i,\delta y_i : i = 1, 2, \dots , N$) pertain to estimates for one physical quantity and that the modelling is appropriate, there may be several 
reasons why the $\chi^2_{\rm min}/{\rm NDF}$ value would exceed $1$.
\begin{itemize}
\item Large(r than statistically expected) random fluctuations in the raw experimental data.
\item Flawed raw experimental data.
\item Invalid assumptions in the analysis of the raw experimental data.
\item Model dependence of the results of the analysis of the raw experimental data; included here are also any cases of ambiguity in the definition of the relevant physical quantities. For the sake of example, the estimates 
for the masses of the resonances are known to be model-dependent: one may define the mass as corresponding to the poles of the $T$- or of the $K$-matrix; alternatively, the mass may be defined as the physical quantity appearing 
in the relativistic Breit-Wigner distribution. As Workman commented in 1999 \cite{Workman1999}: ``the proper way to define and extract the mass of an unstable state continues to be controversial'' and ``the difference between 
Breit-Wigner and pole masses can be $50$ MeV or more.''
\end{itemize}
Such effects have an impact both on the values $y_i$ (which is evident) as well as on their uncertainties $\delta y_i$. In my opinion, however, the dominant reason for what I would call `difficulties in simple fits' is the 
\emph{underestimation of the uncertainties} $\delta y_i$, usually of the systematic ones~\footnote{In a competitive environment, where the esteem by one's colleagues is inversely proportional to the uncertainties in one's 
reported results, the tendency towards the underestimation of the uncertainties, however irresponsible when intentional, might be comprehensible.}: one of the most frequent problems in Physics studies is the underestimation 
of the $\delta y_i$'s in connection with the \emph{model dependence} of the results.

A large $\chi^2$ contribution may be due to two reasons (or their combination): a discrepant input $y_i$ value or an underestimated uncertainty $\delta y_i$. The `usual suspect' in the former case is the model dependence of 
the methodology leading from the raw experimental data to the $y_i$ value. More often than not, the model-dependent effects are of a systematic nature, i.e., causing the systematic under/overestimation of the $y_i$ results. 
To be able to determine the correction which would eliminate such effects, one would need to also perform the analysis in a model-independent way. However, if a model-independent way of extracting the essential information 
were available, one would hardly have opted for a model-dependent analysis in the first place. Therefore, it can be argued that the corrections, removing the model-dependent effects from each $y_i$ value, are unknown (or 
poorly known). As a result, it appears that, in most cases, the practical way to take the model-dependent effects into (some) account, indirectly and in a `handy' manner, is via the enhancement of the uncertainties $\delta y_i$, 
frequently following comparisons of results from two (or even more) model-dependent approaches. One thus comes to compensate for the inability to account for an unknown (or poorly known) bias in the quantities $y_i$, by 
enhancing the corresponding uncertainties $\delta y_i$.

In consequence, given that the model dependence of the input values $y_i$ is frequently hard to assess, one comes, in practice, to attribute the failure of the condition $\chi^2_{\rm min}/{\rm NDF} \approx 1$ to the 
underestimation of the uncertainties $\delta y_i$ \emph{on average}. In such a case, one might opt for an assessment of the `true' uncertainties $\delta y^\prime_i$, which (when used along with the input values $y_i$) would 
bring the $\chi^2_{\rm min}$ result of the fit to the statistical expectation $\chi^2_{\rm min}/{\rm NDF} = 1$. In order that this be achieved, $\delta y_i \to \delta y^\prime_i = s \, \delta y_i$, where $s$ is the Birge 
factor of Eq.~(\ref{eq:EQ07}). This replacement redefines the weight of each input datapoint $w_i \to w^\prime_i = s^{-2} w_i$, leaves the $\avg{y}$ of Eq.~(\ref{eq:EQ02}) intact, but modifies Eq.~(\ref{eq:EQ04}) into
\begin{equation} \label{eq:EQ08}
\delta \avg{y} = \frac{s}{\sqrt{\sum_{i=1}^{N} w_i}} \, \, \, .
\end{equation}
(If all weights $w_i$ are equal to $1$, then $s^2$ is simply the unbiased variance of $N$ observations, and the uncertainty $\delta \avg{y}$ of Eq.~(\ref{eq:EQ08}) is the standard error of the means.)

The Birge factor is applied to the fitted uncertainties (i.e., either to the uncertainties $\delta \avg{y}$ of Eq.~(\ref{eq:EQ04}) or to the MINOS results in case that the MINUIT software package is used in the optimisation) 
only when $s>1$. It is customary to refrain from applying any corrections when the description of the input data comes out `better' than the statistical expectation (i.e., when $s<1$). Therefore, the fitted uncertainties are 
corrected only when they come out (of a fit) too `optimistic'.

To summarise, the weighted averaging of $N$ independent observations $y_i$ with uncertainties $\delta y_i$, the term by which the approach of this section is known, leads to the $\avg{y}$ result of Eq.~(\ref{eq:EQ02}). If 
the fit quality appears to be superior to the statistical expectation, the associated uncertainty $\delta \avg{y}$ is taken from Eq.~(\ref{eq:EQ04}); on the contrary, if the description of the dataset results in $s>1$, then 
the Birge factor is applied to the associated uncertainty $\delta \avg{y}$ of Eq.~(\ref{eq:EQ04}), thus leading to Eq.~(\ref{eq:EQ08}). The approach is straightforward to implement, and is regarded safe even when the input 
dataset contains discrepant observations (outliers), provided that they are \emph{symmetrically distributed} about the $\avg{y}$ value. However, this statistical method yields unreliable results if the outliers are 
\emph{one-sided}, i.e., mostly comprising right/left-tail events in the relevant PDF.

\subsection{\label{sec:MethodsII}Towards robust optimisation}

There are three ways to deal with the presence of outliers in a dataset:
\begin{itemize}
\item[a)] Identification and removal of these datapoints prior to the application of the standard $\chi^2$ method.
\item[b)] Application of weights to these datapoints, leading to reduced contributions to the $\chi^2$ of Eq.~(\ref{eq:EQ01}).
\item[c)] Use of MFs which reduce the contributions from the outliers in a `natural' way. One such example is the logarithmic fit, one example of which is the family of the MFs
\begin{equation*}
{\rm MF} = \frac{1}{k} \sum_{i=1}^{N} \ln \left( 1 + k \, w_i \, (y_i - \avg{y})^2 \right) \, \, \, ,
\end{equation*}
where the real variable $k>0$ is adjustable (usually per category of problems).
\end{itemize}
The removal of the outliers in case (a) may be thought of as an `application of hard weights', whereas the decrease in their contributions in cases (b-c) as due to an `application of soft weights'. In cases (b-c), which may 
in fact be considered as comprising one category, the contributions from the outliers to the MF are smaller than they would have been, had the standard $\chi^2$ MF been used. The more robust a statistical method is, the less 
sensitive its results are expected to be to the presence of outliers in the input dataset.

Investigating the description of an input dataset ($y_i,\delta y_i : i = 1, 2, \dots , N$) by one constant, Ref.~\cite{Matsinos2019} applied seven continuous soft-weight robust methods to a dataset which is known to contain 
outliers. Each of these methods features a tuning parameter $k \in \mathbb{R}^*_+$. Although an adjustable scale, Statistics provides default values for $k$, obtained on the basis of certain assumptions about the distribution 
of the normalised residuals $r_i \coloneqq (y_i - \avg{y})/\delta y_i$. The weights, applied to the input datapoints, aim at reducing the contributions of the elements of the input dataset which give rise to large 
$\lvert r_i \rvert$ values. For convenience, a new variable will be introduced, involving the default value of the tuning parameter $k$ of each method: $z_i=r_i/k$; the weights may thus be thought of as functions of $z_i$, 
rather than of $r_i$. In all cases, the weight in these optimisations (detailed below in alphabetical order) is equal to $1$ for vanishing $z_i$. For non-zero values of the residuals ($z_i \neq 0$), the weights are set as 
follows:
\begin{itemize}
\item Andrews (constant contribution to the MF for large $\lvert r_i \rvert$); default value of the tuning parameter $k=1.339$
\begin{equation} \label{eq:EQ09_01}
W_i (z_i) = \left\{
\begin{array}{rl}
2 z_i^{-2} \left( 1 - \cos \left( z_i \right) \right) & \text{, if $\lvert z_i \rvert < \pi$}\\
4 z_i^{-2} & \text{, otherwise}\\
\end{array} \right.
\end{equation}
\item Cauchy; default value of the tuning parameter $k=2.385$
\begin{equation} \label{eq:EQ09_02}
W_i (z_i) = z_i^{-2}\ln \left( 1 + z_i^2 \right)
\end{equation}
\item Fair; default value of the tuning parameter $k=1.400$
\begin{equation} \label{eq:EQ09_03}
W_i (z_i) = 2 z_i^{-2}\left( \lvert z_i \rvert - \ln \left( 1 + \lvert z_i \rvert \right) \right)
\end{equation}
\item Huber; default value of the tuning parameter $k=1.345$
\begin{equation} \label{eq:EQ09_04}
W_i (z_i) = \left\{
\begin{array}{rl}
1 & \text{, if $\lvert z_i \rvert < 1$}\\
z_i^{-2} \left( 2 \lvert z_i \rvert - 1 \right) & \text{, otherwise}\\
\end{array} \right.
\end{equation}
\item Logistic; default value of the tuning parameter $k=1.205$
\begin{equation} \label{eq:EQ09_05}
W_i (z_i) = 2 z_i^{-2} \ln \left( \cosh \left( z_i \right) \right)
\end{equation}
\item Tukey (constant contribution to the MF for large $\lvert r_i \rvert$); default value of the tuning parameter $k=4.685$
\begin{equation} \label{eq:EQ09_06}
W_i (z_i) = \left\{
\begin{array}{rl}
(3 z_i^2)^{-1} \left( 1 - \left( 1 - z_i^2 \right)^3 \right) & \text{, if $\lvert z_i \rvert < 1$}\\
(3 z_i^2)^{-1} & \text{, otherwise}\\
\end{array} \right.
\end{equation}
\item Welsch; default value of the tuning parameter $k=2.985$
\begin{equation} \label{eq:EQ09_07}
W_i (z_i) = z_i^{-2} \left( 1 - \exp(- z_i^2) \right)
\end{equation}
\end{itemize}
In all cases, the aforementioned weight functions, which are continuous $\forall z_i \in \mathbb{R}$, guarantee that the corresponding seven MFs follow the standard $\chi^2$ MF for small $\lvert r_i \rvert$ values. On the 
other hand, compared to the standard $\chi^2$ case, the relevant contributions to the MF are reduced at large $\lvert r_i \rvert$.

The most robust of the aforementioned methods, hence the least sensitive to the presence of outliers, are the methods of Andrews and Tukey: in both cases, the contributions from the outliers are constant (i.e., independent 
of the distance of such datapoints to $\avg{y}$). Although the results are expected to be similar when using these two statistical methods, both fits will be performed (and their results will be reported) in all cases: this 
was deemed important for the verification of the results and for completeness.

It would be interesting to have a closer look at the definition of datapoints as outliers in these two robust-optimisation methods. In case of Andrews weights, the outliers are defined as the datapoints which satisfy the 
condition $\lvert z_i \rvert \geq \pi$, which implies that $\lvert r_i \rvert \geq \pi k$ or $\lvert r_i \rvert \gtrapprox 4.207$ for the default value for $k$; in case of Tukey weights, the outliers are defined as the 
datapoints which satisfy the condition $\lvert z_i \rvert \geq 1$, which implies that $\lvert r_i \rvert \geq k$ or $\lvert r_i \rvert \gtrapprox 4.685$ for the default value for $k$. Evidently, the former method is expected 
to lead to the identification of more outliers. There is no doubt that both methods lead to the identification of the so-called distant (or obvious) outliers, leaving in the original sample datapoints which make large 
contributions to the standard $\chi^2$ MF; this must be borne in mind when interpreting the fit results in Section \ref{sec:Results}. To exclude more datapoints, one could opt to use lower $k$ values than the default ones. 
However, I will refrain from applying changes to established statistical methods herein.

To summarise, several approaches are available for handling datasets which contain outliers, including the use in the standard $\chi^2$ MF of the seven types of weights of Eqs.~(\ref{eq:EQ09_01}-\ref{eq:EQ09_07}). Apart from 
extracting reliable estimates from the input data, such methods may evidently be used as a means of identifying the outliers in the input dataset. The important advantage of the robust-optimisation schemes is the dynamical 
identification of outliers: the outliers are present in the sample at all times, and only the input data themselves decide which datapoint should be an outlier and which not (at any given step of the optimisation). The 
robust-optimisation methods have been developed in order to enable the optimal description of the \emph{bulk} of the input datapoints. If the bulk of a dataset comprises unreliable measurements, then nothing can be done: I 
am not aware of data-driven statistical methods which could yield reliable results from an unreliable core of input data.

\subsection{\label{sec:MethodsIII}Identification of the outliers}

The identification of the outliers is straightforward when using the methods of Andrews, Huber, and Tukey: regular datapoints and outliers are clearly distinguished (at any given step of the optimisation) on the basis of 
their $r_i$ value. The method of Huber accepts linear contributions from the outliers (as opposed to the quadratic contributions to the standard $\chi^2$ MF); although useful in several situations, the emphasis herein is 
placed on minimal sensitivity to the outliers.

It should be mentioned that dedicated algorithms for the identification of the outliers in datasets have been available since the late 1960s (e.g., see Refs.~\cite{Grubbs1969,Stefansky1972,Rosner1983}, yet they do not 
accommodate the input uncertainties $\delta y_i$. There are some difficulties in coming up with a variant of the existing algorithms \cite{Grubbs1969,Stefansky1972,Rosner1983} while also accounting for individual weights. 
However, there is another option for those who favour the use of conventional methods in their analyses: one may put forward a simple $\chi^2$-based iterative procedure as follows. One performs the $\chi^2$ fit using the 
original dataset (which might contain outliers) and, provided that the resulting p-value of that fit satisfies the condition ${\rm p} < {\rm p}_{\rm min}$, one examines the fitted results for the normalised residuals, 
identifies the datapoint $\arg \max \{ \lvert r_i \rvert : i = 1, 2, \dots , N \}$, and removes it from the input dataset in the subsequent fit. One repeats this procedure until ${\rm p} \geq {\rm p}_{\rm min}$. The 
resulting subset is devoid of outliers and may be identified as the bulk of the original dataset (at significance level ${\rm p}_{\rm min}$). This procedure is an acceptable alternative to the use of robust techniques in 
the optimisation.

\subsection{\label{sec:MethodsIV}Results obtained on the basis of the CDF}

This method of providing estimates for the quantities $\avg{y}$ and $\delta \avg{y}$ was applied in Ref.\cite{Matsinos2019} to results exhibiting sizeable variability. Given an input dataset ($y_i,\delta y_i : i = 1, 2, \dots , N$), 
one may obtain the CDF of the quantity which is being studied by averaging the CDFs obtained from all input datapoints. This non-parametric determination rests upon the assumption that each input datapoint $(y_i,\delta y_i)$ 
gives rise to a Gaussian distribution with average $\mu_i=y_i$ and variance $\sigma^2_i=(\delta y_i)^2$. The median of the resulting CDF is taken to be a meaningful estimate for $\avg{y}$, whereas the $1 \sigma$ limits (below 
and above the median value) are proposed as estimates for the negative and positive root-mean-square (rms) values of the distribution, which are associated with the standard errors (negative and positive) of the means. 
Although the method appears to be robust enough to be employed even in the presence of (a reasonable amount of) outliers, it will be applied herein to datasets devoid of (distant) outliers. Technically, the CDF at point $x$ 
is obtained using the formula
\begin{equation} \label{eq:EQ10}
{\rm CDF} (x) = \frac{1}{N} \sum_{i=1}^{N} \Phi \left( \frac{x-y_i}{\delta y_i} \right) \, \, \, ,
\end{equation}
where $\Phi(x)$ is the normal CDF. The asymmetrical uncertainties are easily accommodated in this scheme on account of the sign of each quantity $x-y_i$.

One of the interesting features of this method is that it does not apply any weights $w_i$ to the input datapoints; the magnitude of each input uncertainty $\delta y_i$ manifests itself in the size of the width of the 
distribution contributed to the CDF. An additional advantage is that the method makes use of the median value, which is generally considered to be more representative (of the input dataset) than the average.

\section{\label{sec:Results}Results}

Before entering the details, I will list the steps which are assumed in the analysis of each input dataset ($y_i,\delta y_i : i = 1, 2, \dots , N$), corresponding to estimates for one physical quantity.
\begin{itemize}
\item First, the description of the dataset will be attempted using the robust-optimisation methods of Andrews and Tukey. The extracted values with these two statistical methods are the \emph{main results} of this study. 
Consequently, the effects of the outliers, albeit in a reduced form according to the weights of Eqs.~(\ref{eq:EQ09_01}) and (\ref{eq:EQ09_06}), are also contained in the recommended values of this work.
\item The datapoints comprising the union of the two sets of outliers, obtained from the output of the two aforementioned robust methods, will be removed from the dataset, to yield the so-called `trimmed' dataset, taken in 
this work to be the bulk of the original input data. Evidently, the trimmed and the original datasets are identical if no outliers can be found.
\item The description of the trimmed dataset will be investigated using the standard $\chi^2$ MF.
\item The trimmed dataset will be used in the determination of the CDF, as explained in Section \ref{sec:MethodsIV}.
\item The estimates obtained from the trimmed dataset will be considered as \emph{supplementary}, serving the purpose of testing the self-consistency of this study, which dictates that all four results be compatible among 
themselves.
\end{itemize}

The exclusive source of input herein is Ref.~\cite{pdg2020}; no attempt was made towards the verification of the correctness of the values listed therein. Values, which are not accompanied by an uncertainty, were omitted. 
The left-side (negative) uncertainties, statistical and systematic (wherever available), will be summed in quadrature, to yield one negative uncertainty; similarly, a positive uncertainty will be derived. Both uncertainties 
will be taken into account in the optimisation; the modifications, required in order that the asymmetrical uncertainties be accommodated in the optimisation, are straightforward to implement (on the basis of the sign of each 
normalised residual). One should mention that the MINUIT method MINOS evaluates and outputs (by default) the asymmetrical fitted uncertainties for each fit parameter. If, after their truncation to two significant digits, the 
asymmetrical uncertainties do not come out equal, both uncertainties will be given.

At this point, one word of caution is due. The main weakness of this work appertains to the use of all data in the ten relevant particle listings of Ref.~\cite{pdg2020} as independent observations. Due to three reasons, not 
all listed values are independent.
\begin{itemize}
\item Some values might have resulted from analyses of identical or overlapping databases.
\item Some values appear to be updates of earlier estimates (e.g., results of works which might have applied improved methodologies to identical or overlapping databases).
\item Some values have been taken from compilations of physical constants: such estimates are not independent from the observations they have been based on.
\end{itemize}
As it would have entailed extensive time investment (one would have to get hold of and pore over hundreds of papers), no attempt was made in this work towards the validation of the input (regarding the aforementioned three 
sources of bias). On the other hand, this information is surely available to the PDG, who could decide to make it available to all. Incidentally, a more convenient categorisation of the data appearing in the PDG compilations 
(e.g., independent observations satisfying their selection criteria, superseded results, problematic data, etc.) would be helpful to all researchers who intend to submit the available data to different processing. Equally 
welcomed would be improvements regarding the functionality of their webpage, for instance, to enable the retrieval of the data in the form of simple (e.g., ASCII) files.

\subsection{\label{sec:ScalarIsoscalar}Scalar-isoscalar mesons}

This section relates to the properties of the scalar-isoscalar mesons. The summary of the recommended values (also including those favoured by the PDG) can be found in Table \ref{tab:ScalarIsoscalar}. Details about the data 
analysis can be found in Sections \ref{sec:f0500}-\ref{sec:f01710}.

Among the decay modes of the scalar-isoscalar mesons are those to the $\pi^+ \pi^-$ and $\pi^0 \pi^0$ final states. Both decays conserve parity and G-parity. The Clebsch-Gordan coefficients $\bra{I_1, I_2 ; I_{1z}, I_{2z}}\ket{I ; I_z}$, 
where $I=I_z=0$ ($f_0$ in the initial state), and
\begin{itemize}
\item $I_1=I_2=I_{1z}=1$, $I_{2z}=-1$ ($\pi^+ \pi^-$ in the final state) and 
\item $I_1=I_2=1$, $I_{1z}=I_{2z}=0$ (two $\pi^0$'s in the final state)
\end{itemize}
are equal (but of opposite sign), thus suggesting equal branching fractions. To come up with meaningful predictions for the two branching fractions, one would need to consider the available phase space in each case.

\begin{table}%[h!]
{\bf \caption{\label{tab:ScalarIsoscalar}}}Summary of the recommended values for the physical properties of the scalar-isoscalar mesons below $2$ GeV. The masses and the total decay widths are expressed in MeV, the branching 
fractions in percent. The PDG notation regarding the method yielding their results is as follows. AVERAGE: from a weighted average of selected data. ESTIMATE: based on the observed range of the data; not from a formal 
statistical procedure. EVALUATION: not from a direct measurement, but evaluated from measurements of related quantities. FIT: from a constrained or overdetermined multiparameter fit of selected data.
\vspace{0.3cm}
\begin{center}
\begin{tabular}{|c|c|c|}
\hline
Physical quantity & PDG \cite{pdg2020} & This work\\
\hline
\hline
\multicolumn{3}{|c|}{$f_0(500)$}\\
\hline
Mass & $400-550$ (ESTIMATE) & $497^{+28}_{-33}$\\
Total decay width & $400-700$ (ESTIMATE) & $381^{+74}_{-75}$\\
Branching fraction of $f_0(500) \to \pi \pi$ & $\approx 100$ & $\approx 100$\\
\hline
\multicolumn{3}{|c|}{$f_0(980)$}\\
\hline
Mass & $990 \pm 20$ (ESTIMATE) & $979.4^{+4.0}_{-3.8}$\\
Total decay width & $10-100$ (ESTIMATE) & $49.2^{+22.6}_{-5.1}$\\
Branching fraction of $f_0(980) \to \pi \pi$ & $-$ & $81.0^{+3.0}_{-3.2}$\\
\hline
\multicolumn{3}{|c|}{$f_0(1370)$}\\
\hline
Mass & $1200-1500$ (ESTIMATE) & $1320 \pm 12$\\
Total decay width & $200-500$ (ESTIMATE) & $233^{+44}_{-54}$\\
Branching fraction of $f_0(1370) \to \pi \pi$ & $-$ & $-$\\
\hline
\multicolumn{3}{|c|}{$f_0(1500)$}\\
\hline
Mass & $1506 \pm 6$ (AVERAGE) & $1498.5^{+4.4}_{-4.2}$\\
Total decay width & $112 \pm 9$ (AVERAGE) & $117.2 \pm 4.1$\\
Branching fraction of $f_0(1500) \to \pi \pi$ & $34.5 \pm 2.2$ (FIT) & $28.2^{+5.3}_{-2.8}$\\
\hline
\multicolumn{3}{|c|}{$f_0(1710)$}\\
\hline
Mass & $1732^{+9}_{-7}$ (AVERAGE) & $1715.1^{+10.5}_{-9.2}$\\
 & $1704 \pm 12$ (EVALUATION) & $-$\\
Total decay width & $147^{+12}_{-10}$ (AVERAGE) & $144 \pm 13$\\
 & $123 \pm 18$ (EVALUATION) & $-$\\
Branching fraction of $f_0(1710) \to \pi \pi$ & $-$ & $8.9^{+3.2}_{-3.0}$\\
\hline
\end{tabular}
\end{center}
\vspace{0.5cm}
\end{table}

\subsubsection{\label{sec:f0500}$f_0(500)$}

Regarding the scalar-isoscalar mesons, this is the most important of the $t$-channel exchanges in the $\pi N$ interaction model of Ref.~\cite{matsinos2017}.

Under the Sections `$f_0(500)$ $T$-matrix pole $\sqrt{s}$' and `$f_0(500)$ Breit-Wigner mass or $K$-matrix pole parameters', the PDG list a number of results for the mass of the $f_0(500)$, which are nevertheless preceded by 
the remark: ``We do not use the following data for averages, fits, limits, etc.'' The PDG opt for a rough guess as to the significant domain of the PDF of the mass of the $f_0(500)$, and recommend the use of the range from 
$400$ to $550$ MeV.

I decided to accept all $42$ listed values as comprising independent observations and submit them to the optimisation using the statistical methods of Andrews and Tukey. The Birge factor of these fits came out large, about 
$2.50$, indicating discrepant input (thus confirming the overall impression, obtained from the visual inspection of the plot of the original input datapoints). In both cases, $\avg{y}$ came out about $497$ MeV, with a fitted 
uncertainty of about $30$ MeV, see Table \ref{tab:f0500}, upper part.

\begin{table}%[h!]
{\bf \caption{\label{tab:f0500}}}Physical properties of the $f_0(500)$. The statistical methods using Andrews and Tukey weights are robust and admit the entire set of available results. Regarding the methods featuring the 
$\chi^2$ MF and the CDF, eleven outliers in case of the mass and four in case of the total decay width, identified as such by the two aforementioned robust methods, were removed from the original input dataset.
\vspace{0.3cm}
\begin{center}
\begin{tabular}{|c|c|}
\hline
Source/Method & Result\\
\hline
\hline
\multicolumn{2}{|c|}{Mass (MeV)}\\
\hline
PDG \cite{pdg2020} & $400-550$\\
Andrews weights & $497^{+28}_{-33}$\\
Tukey weights & $497^{+27}_{-32}$\\
Standard $\chi^2$ & $489.1 \pm 7.8$\\
CDF & $498.8^{+12.3}_{-9.5}$\\
\hline
\multicolumn{2}{|c|}{Total decay width (MeV)}\\
\hline
PDG \cite{pdg2020} & $400-700$\\
Andrews weights & $381^{+74}_{-75}$\\
Tukey weights & $381^{+75}_{-76}$\\
Standard $\chi^2$ & $387 \pm 28$\\
CDF & $374^{+40}_{-32}$\\
\hline
\multicolumn{2}{|c|}{Branching fraction of $f_0(500) \to \pi \pi$ ($\%$)}\\
\hline
PDG \cite{pdg2020} & $\approx 100$\\
This work & $\approx 100$\\
\hline
\end{tabular}
\end{center}
\vspace{0.5cm}
\end{table}

\begin{figure}
\begin{center}
\includegraphics [width=15.5cm] {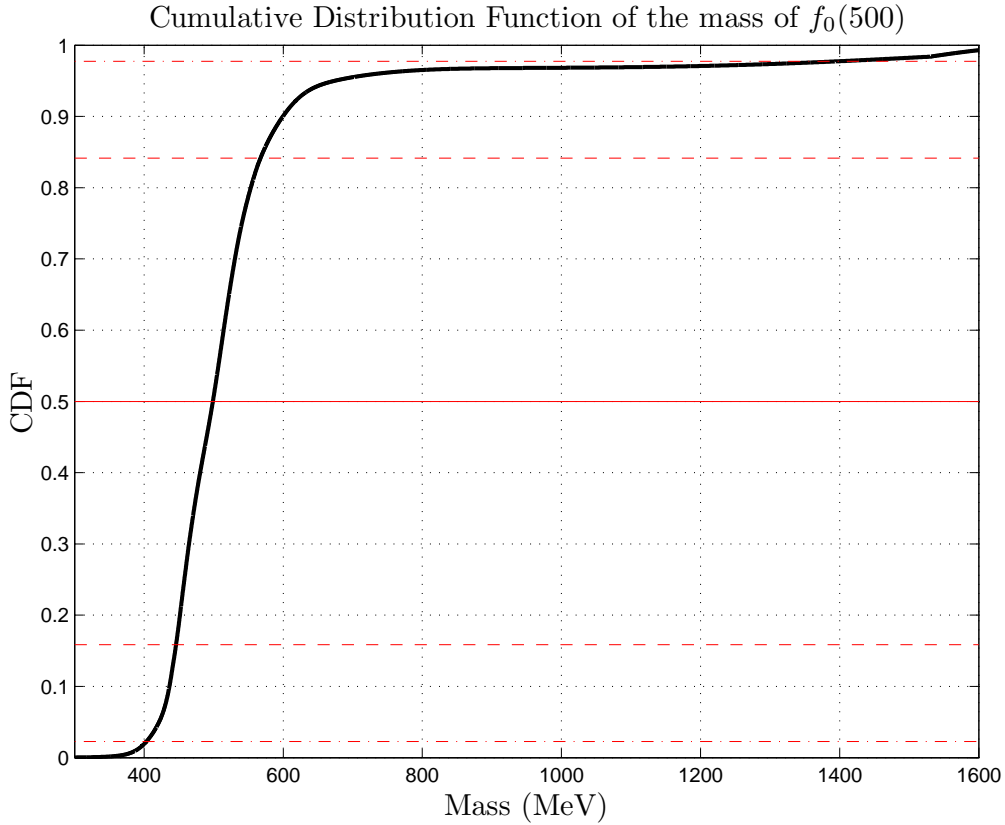}
\caption{\label{fig:MassOff0500}The CDF of the mass of the $f_0(500)$, obtained from averaging $31$ Gaussian distributions, see Eq.~(\ref{eq:EQ10}). Eleven outliers were removed on the basis of the results of the robust 
methods of Andrews and Tukey (see Section \ref{sec:MethodsIII}). The horizontal solid straight line marks the $50~\%$ level; the two horizontal dashed straight lines delineate the $68.27~\%$ Confidence Interval (CI), the 
equivalent of $1 \sigma$ limits in the normal distribution; the two horizontal dashed-dotted straight lines delineate the $95.45~\%$ CI, the equivalent of $2 \sigma$ limits in the normal distribution.}
\vspace{0.35cm}
\end{center}
\end{figure}

The range of the $42$ average values of the available estimates extends from $414$ to $1530$ MeV. The two robust methods identified eleven outliers, labelled in Ref.~\cite{pdg2020} as MOUSSALLAM11, PELAEZ04A, SUROVTSEV01, and 
ESTABROOKS79 in their Section `$f_0(500)$ $T$-matrix pole $\sqrt{s}$', and those labelled as ALEKSEEV99 \& 98, TROYAN98, ALDE97, ISHIDA97, SVEC96, and ANISOVICH95 in their Section `$f_0(500)$ Breit-Wigner mass or $K$-matrix 
pole parameters'. The fit to the $31$ remaining datapoints with the standard $\chi^2$ MF yielded a Birge factor of about $2.07$ and a fitted $\avg{y}$ value compatible with (though slightly lower than) the results of the two 
robust methods. The CDF, obtained from the same $31$ datapoints, is displayed in Fig.~\ref{fig:MassOff0500}. The $50~\%$ level of the CDF yields the median value of the distribution, which is in good agreement with the 
results of the two robust methods.

There are two reasons why the fitted uncertainties come out larger in the robust fits (in comparison with the standard $\chi^2$ fits): a) The weights $W_i$ of Eqs.~(\ref{eq:EQ09_01}-\ref{eq:EQ09_07}) satisfy $W_i \leq 1$. This 
leads to an effective (positive real number) ${\rm NDF} \coloneqq \sum_{i=1}^{N} W_i - 1$ below the (positive integer) NDF of the corresponding $\chi^2$ fit (where $W_i \equiv 1$), even though no outliers are present in the 
latter case (hence they do not contribute to the NDF). b) More often than not, the $\chi^2_{\rm min}$ values come out larger in the robust fits.

Twelve results on the total decay width of the $f_0(500)$ are found in Ref.~\cite{pdg2020}. They range between $35(12)$ (TROYAN98) and $780(60)$ (ALDE97) MeV, making one wonder whether such a spread of values could refer 
to determinations of one physical quantity. The application of the approach, put forward in the beginning of Section \ref{sec:Results}, yielded the results of Table \ref{tab:f0500}, middle part. As expected, the estimates, 
obtained with the two robust methods, agree well. Four datapoints (the data labelled in Ref.~\cite{pdg2020} as ALEKSEEV99 \& 98, TROYAN98, and ALDE97) were identified as outliers. The results of the standard $\chi^2$ method, 
as well as of the analysis featuring the determination of the properties of the resonances from the CDF, corroborate the estimates of the robust optimisation. The resulting p-value of the $\chi^2$ fit is equal to about 
$8.81 \cdot 10^{-2}$, indicating an acceptable fit.

\begin{figure}
\begin{center}
\includegraphics [width=15.5cm] {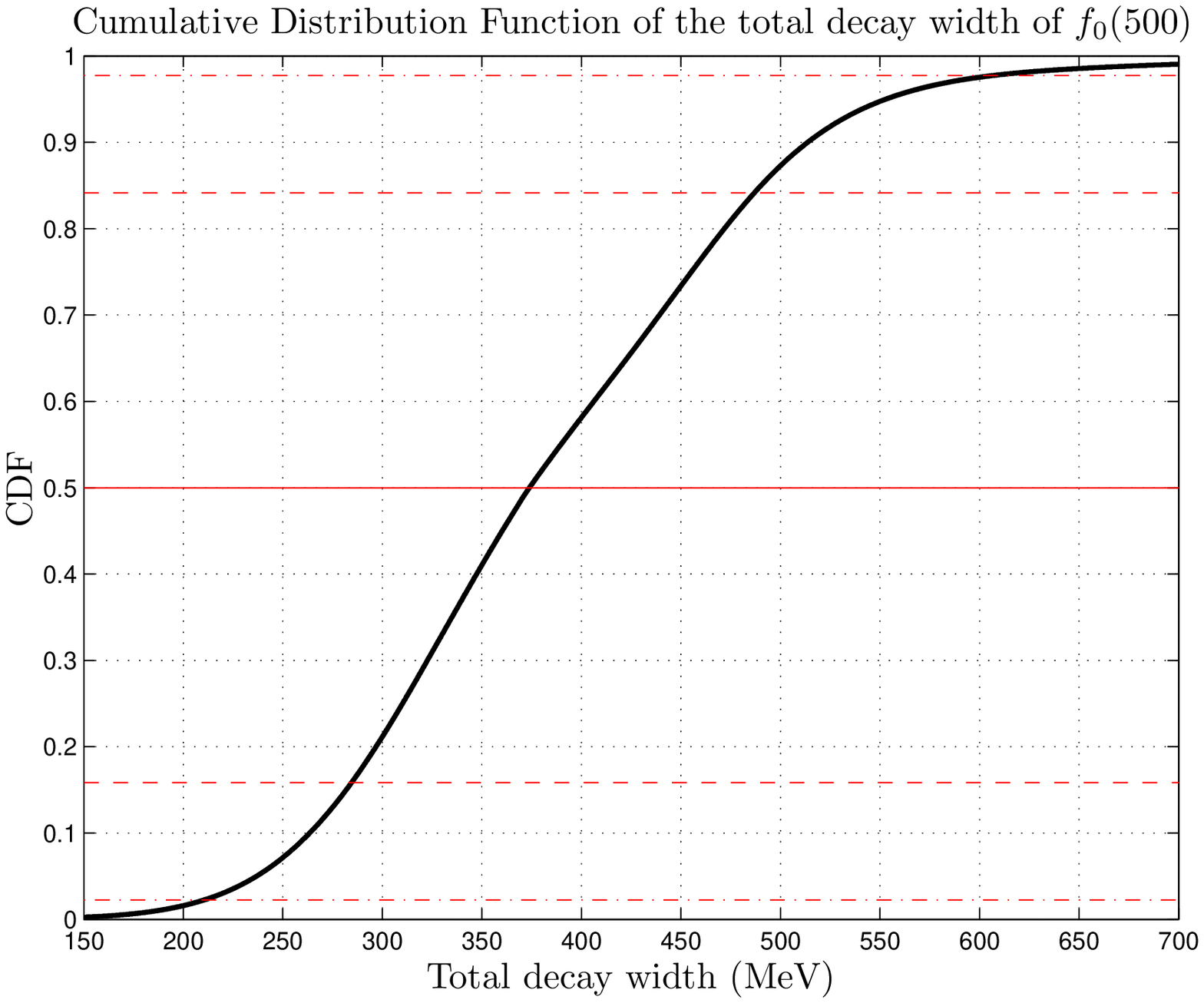}
\caption{\label{fig:WidthOff0500}The CDF of the total decay width of the $f_0(500)$ obtained from averaging eight Gaussian distributions. The horizontal straight lines represent the same levels as in Fig.~\ref{fig:MassOff0500}.}
\vspace{0.35cm}
\end{center}
\end{figure}

Reference \cite{pdg2020} lists two decay modes of the $f_0(500)$: to $\pi \pi$ and to $\gamma \gamma$. Although both modes are marked as `seen', inspection of the data for the partial decay width $f_0(500) \to \gamma \gamma$ 
reveals that this mode can safely be omitted. The application of the two robust methods resulted in $\Gamma (f_0(500) \to \gamma \gamma) \approx 1.95$ keV (with an uncertainty of about $0.20$ keV), negligible when compared 
to the aforementioned estimates for the total decay width of the $f_0(500)$.

In summary, the analysis of the mass values of the $f_0(500)$, listed in Ref.~\cite{pdg2020}, suggests a mass close to $500$ MeV with a $1 \sigma$ uncertainty of about $30$ MeV. This range of values corresponds to a narrower 
domain than the one recommended by the PDG. The total decay width of the $f_0(500)$ appears to be about $380$ MeV, short of the lower limit of $400$ MeV recommended by the PDG, with an uncertainty of about $75$ MeV. Therefore, 
the width of the window of the most probable values ($1 \sigma$ limits, $68.27~\%$ confidence) is about $150$ MeV, i.e., half the extent of the corresponding domain recommended by the PDG. The branching fraction of the 
two-pion decay mode can safely be taken to be equal to $100~\%$.

\subsubsection{\label{sec:f0980}$f_0(980)$}

\begin{figure}
\begin{center}
\includegraphics [width=15.5cm] {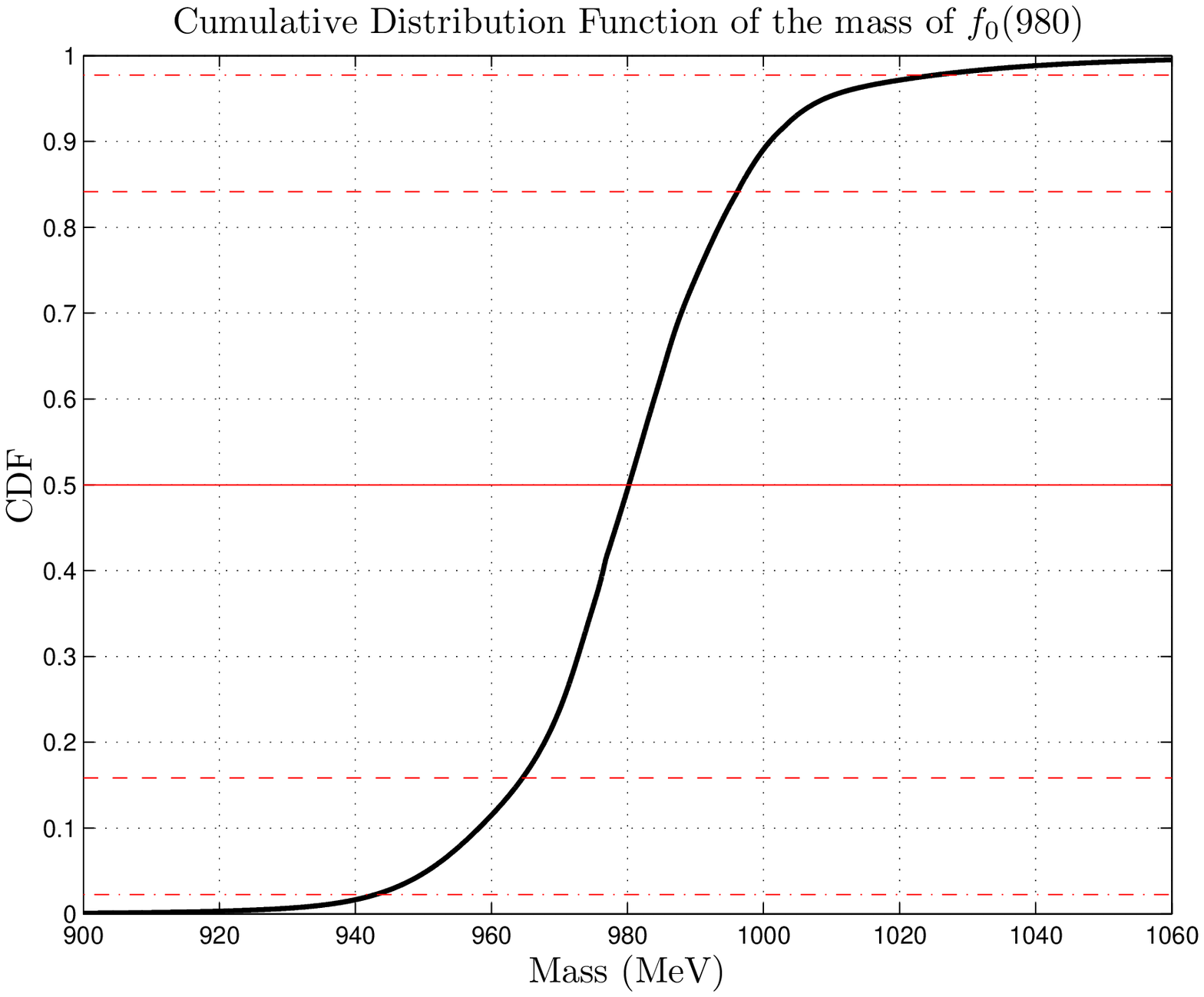}
\caption{\label{fig:MassOff0980}The equivalent of Fig.~\ref{fig:MassOff0500} for the mass of the $f_0(980)$.}
\vspace{0.35cm}
\end{center}
\end{figure}

Regarding the mass of the $f_0(980)$, the PDG recommend the use of $990(20)$ MeV. The application of the two robust methods to the original input dataset of $61$ estimates resulted in the fitted values and uncertainties of 
Table \ref{tab:f0980}, upper part. Both statistical methods identified six outliers, labelled in Ref.~\cite{pdg2020} as ABLIKIM15P \& 12E, ANISOVICH09 \& 03, ALOISIO02D, and GRAYER73. These datapoints were removed from the 
dataset before it was submitted to the optimisation using the standard $\chi^2$ MF, as well as to the analysis featuring the determination of the properties of the resonances from the CDF. In the former case, the Birge 
factor comes out equal to about $1.81$. The CDF of the mass of the $f_0(980)$ is displayed in Fig.~\ref{fig:MassOff0980}.

\begin{table}%[h!]
{\bf \caption{\label{tab:f0980}}}The equivalent of Table \ref{tab:f0500} for the $f_0(980)$. Regarding the methods featuring the $\chi^2$ MF and the CDF, six outliers in case of the mass and five in case of the total decay 
width, identified as such by the robust-optimisation methods of Andrews and Tukey, were removed from the original input dataset.
\vspace{0.3cm}
\begin{center}
\begin{tabular}{|c|c|}
\hline
Source/Method & Result\\
\hline
\hline
\multicolumn{2}{|c|}{Mass (MeV)}\\
\hline
PDG \cite{pdg2020} & $990 \pm 20$\\
Andrews weights & $979.4^{+4.0}_{-3.8}$\\
Tukey weights & $979.5^{+3.6}_{-3.7}$\\
Standard $\chi^2$ & $979.5 \pm 1.3$\\
CDF & $980.2 \pm 2.1$\\
\hline
\multicolumn{2}{|c|}{Total decay width (MeV)}\\
\hline
PDG \cite{pdg2020} & $10-100$\\
Andrews weights & $49.2^{+22.6}_{-5.1}$\\
Tukey weights & $49.1^{+22.0}_{-5.0}$\\
Standard $\chi^2$ & $50.5 \pm 2.8$\\
CDF & $58.3^{+4.7}_{-4.1}$\\
\hline
\multicolumn{2}{|c|}{Branching fraction of $f_0(980) \to \pi \pi$ ($\%$)}\\
\hline
PDG \cite{pdg2020} & $-$\\
Andrews weights & $81.0^{+3.0}_{-3.2}$\\
Tukey weights & $81.0^{+3.0}_{-3.2}$\\
Standard $\chi^2$ & $81.0^{+2.4}_{-2.3}$\\
CDF & $77.2^{+3.0}_{-7.8}$\\
\hline
\end{tabular}
\end{center}
\vspace{0.5cm}
\end{table}

\begin{figure}
\begin{center}
\includegraphics [width=15.5cm] {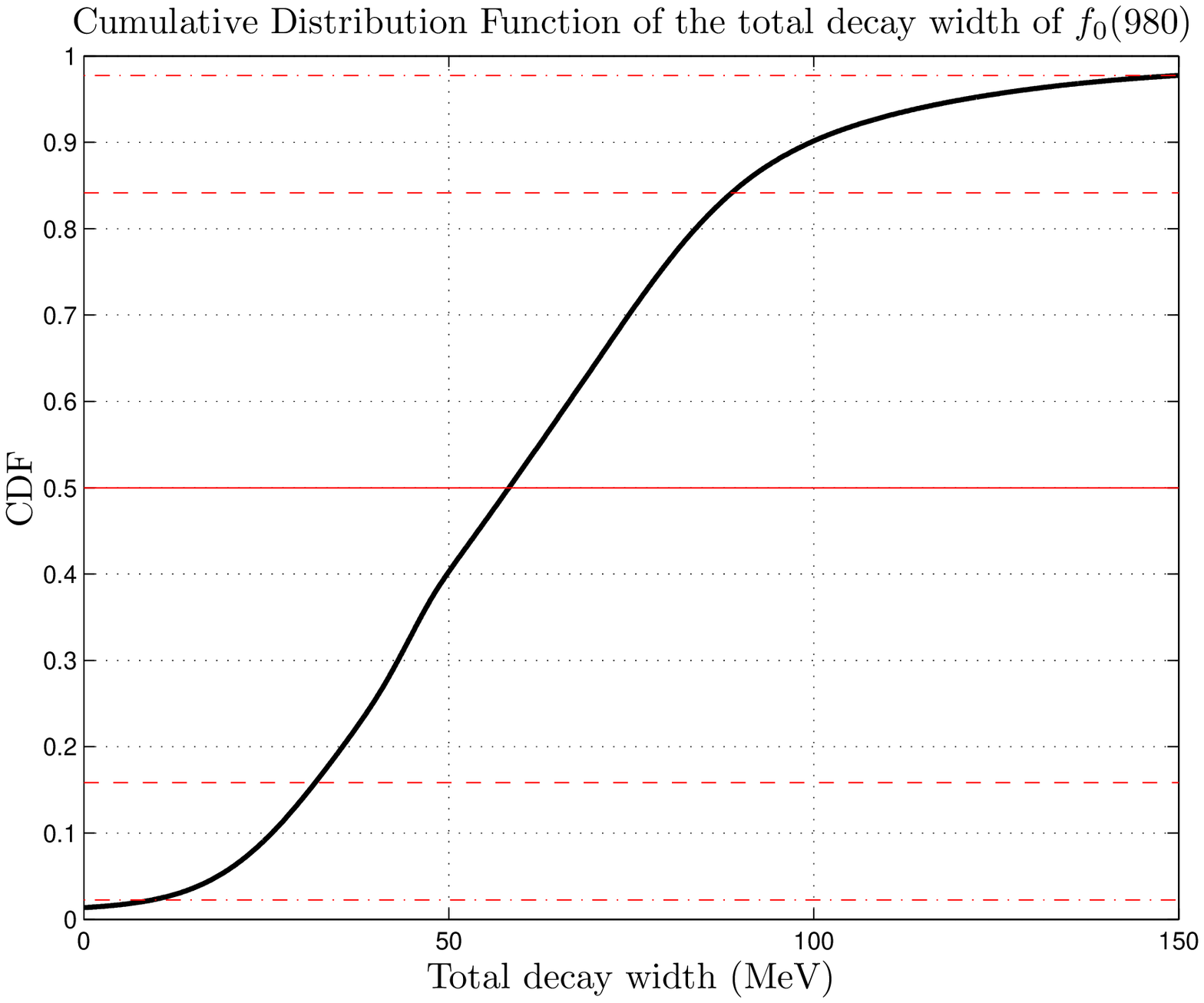}
\caption{\label{fig:WidthOff0980}The equivalent of Fig.~\ref{fig:WidthOff0500} for the total decay width of the $f_0(980)$.}
\vspace{0.35cm}
\end{center}
\end{figure}

Out of the $47$ results for the total decay width of the $f_0(980)$, five were identified as outliers: they are the values labelled in Ref.~\cite{pdg2020} as ABLIKIM15P \& 12E, ACHASOV00H (both estimates), and KAMINSKI94. 
The estimates of this work are contained in Table \ref{tab:f0980}, middle part, and the corresponding CDF is displayed in Fig.~\ref{fig:WidthOff0980}. In case of the standard $\chi^2$ fit, the Birge factor comes out equal to 
about $1.61$. The result of the CDF method slightly exceeds the one obtained with the standard $\chi^2$ MF because of the presence in the trimmed dataset of large values with large uncertainties, e.g., the datapoints 
labelled as TIKHOMIROV03 ($121(23)$ MeV), BREAKSTONE90 ($110(30)$ MeV), ETKIN82B ($120(281)(20)$ MeV), and AGUILAR-BENITEZ78 ($100(80)$ MeV): owing to the largeness of their uncertainties, these one-sided results do not 
emerge as outliers.

The PDG list four decay modes of the $f_0(980)$: to the $\pi \pi$, $K \bar{K}$, $\gamma \gamma$, and $e^+ e^-$ final states, all marked as `seen' in their compilation. The partial decay width to $\gamma \gamma$ is evaluated 
in Ref.~\cite{pdg2020} to a mere $0.31^{+0.05}_{-0.04}$ keV, hence it can safely be omitted; the same goes for the $e^+ e^-$ final state, whose partial decay width is below $8.4$ eV at $90~\%$ confidence. For the ratio 
$\Gamma(\pi \pi) / (\Gamma(\pi \pi) + \Gamma(K \bar{K}))$, representing for all practical purposes the branching fraction of $f_0(980) \to \pi \pi$, the PDG mention six estimates (with uncertainties), but put forward no 
recommendation. The results of this work are listed in Table \ref{tab:f0980}, lower part. None of the six datapoints emerges from the robust fit as outlier; the fit using the $\chi^2$ MF is acceptable, as it yields the 
p-value of about $3.75 \cdot 10^{-2}$, i.e., exceeding the ${\rm p}_{\rm min}$ of this work.

In summary, the recommended value of this work for the mass of the $f_0(980)$ is about $980$ MeV, with an uncertainty of about $4$ MeV, i.e., about $0.5 \sigma$ short of the PDG estimate of $990(20)$ MeV. The recommended 
range of values of this work for the total decay width of the $f_0(980)$ is more restricted compared to the PDG interval from $10$ to $100$ MeV. The data analysis results in a large branching fraction of the $f_0(980)$ 
decay to two pions, about $81~\%$, with an uncertainty of about $3~\%$.

\subsubsection{\label{sec:f01370}$f_0(1370)$}

\begin{figure}
\begin{center}
\includegraphics [width=15.5cm] {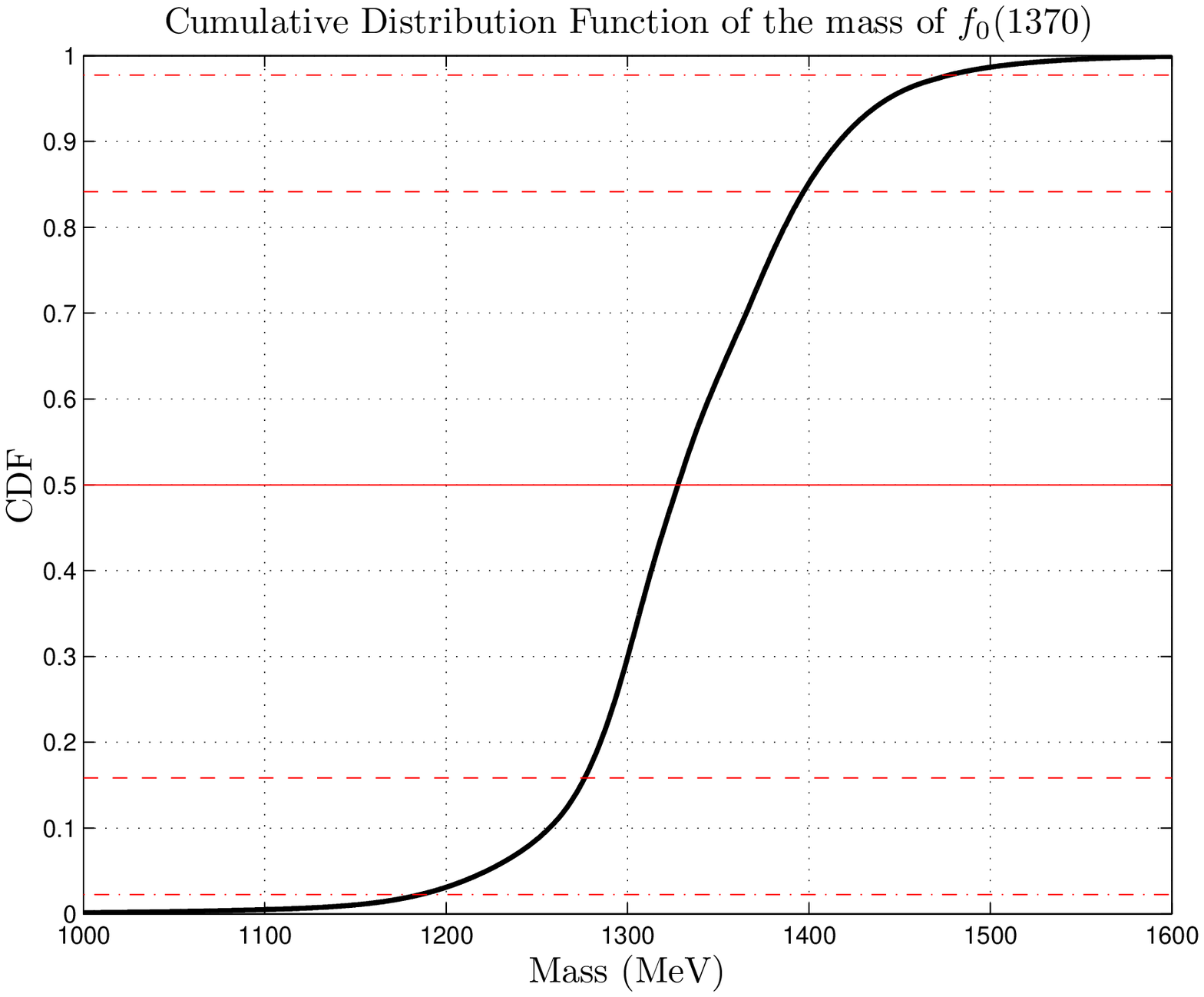}
\caption{\label{fig:MassOff01370}The equivalent of Fig.~\ref{fig:MassOff0500} for the mass of the $f_0(1370)$.}
\vspace{0.35cm}
\end{center}
\end{figure}

As earlier for the $f_0(500)$, the datasets contained in the two sections of Ref.~\cite{pdg2020}, to be specific in `$f_0(1370)$ $T$-matrix pole position' and in `$f_0(1370)$ Breit-Wigner mass or $K$-matrix pole parameter', 
were combined into one set. The application of the two robust methods to the $47$ estimates of the combined set resulted in the fitted values and uncertainties of Table \ref{tab:f01370}, upper part. Eleven datapoints were 
identified as outliers, labelled in Ref.~\cite{pdg2020} as KAMINSKI99, BERTIN97C, and KAMINSKI94 in the former section, and GARMASH06, AITALA01A, ARMSTRONG91, and AKESSON86 (from the $\pi \pi$ mode), VLADIMIRSKY06, TIKHOMIROV03, 
ETKIN82B, and WICKLUND80 (from the $K \bar{K}$ mode) in the latter. These datapoints were removed from the dataset before it was submitted to the optimisation using the standard $\chi^2$ MF, as well as to the analysis 
featuring the determination of the properties of the resonances from the CDF. In the former case, the Birge factor comes out equal to about $1.57$. The CDF of the mass of the $f_0(1370)$ is displayed in Fig.~\ref{fig:MassOff01370}.

\begin{table}%[h!]
{\bf \caption{\label{tab:f01370}}}The equivalent of Table \ref{tab:f0500} for the $f_0(1370)$. Regarding the methods featuring the $\chi^2$ MF and the CDF, eleven outliers in case of the mass and six in case of the total 
decay width, identified as such by the robust-optimisation methods of Andrews and Tukey, were removed from the original input dataset.
\vspace{0.3cm}
\begin{center}
\begin{tabular}{|c|c|}
\hline
Source/Method & Result\\
\hline
\hline
\multicolumn{2}{|c|}{Mass (MeV)}\\
\hline
PDG \cite{pdg2020} & $1200-1500$\\
Andrews weights & $1320 \pm 12$\\
Tukey weights & $1320 \pm 12$\\
Standard $\chi^2$ & $1324.2 \pm 5.7$\\
CDF & $1327.9^{+11.5}_{-8.6}$\\
\hline
\multicolumn{2}{|c|}{Total decay width (MeV)}\\
\hline
PDG \cite{pdg2020} & $200-500$\\
Andrews weights & $233^{+44}_{-54}$\\
Tukey weights & $230^{+51}_{-49}$\\
Standard $\chi^2$ & $233 \pm 15$\\
CDF & $250 \pm 22$\\
\hline
\multicolumn{2}{|c|}{Branching fraction of $f_0(1370) \to \pi \pi$ ($\%$)} \\
\hline
PDG \cite{pdg2020} & $-$\\
This work & $-$\\
\hline
\end{tabular}
\end{center}
\vspace{0.5cm}
\end{table}

\begin{figure}
\begin{center}
\includegraphics [width=15.5cm] {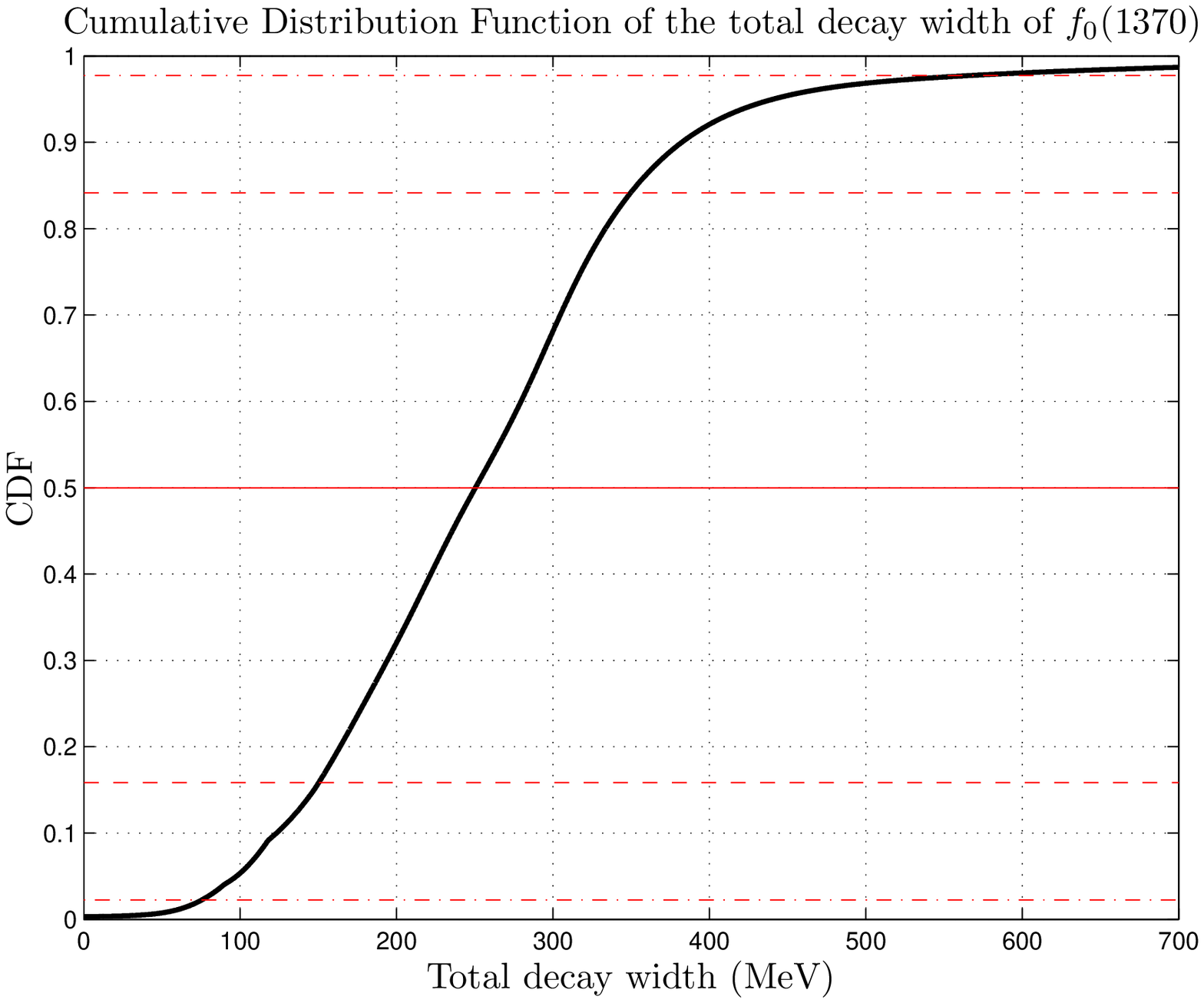}
\caption{\label{fig:WidthOff01370}The equivalent of Fig.~\ref{fig:WidthOff0500} for the total decay width of the $f_0(1370)$.}
\vspace{0.35cm}
\end{center}
\end{figure}

Out of the $27$ results for the total decay width of the $f_0(1370)$ (corresponding to all modes), six were identified as outliers: they are the values labelled in Ref.~\cite{pdg2020} as GARMASH06, BERTIN98, and AKESSON86 
(from the $\pi \pi$ mode), VLADIMIRSKY06 and TIKHOMIROV03 (from the $K \bar{K}$ mode), and ADAMO93 (from the $4 \pi$ mode). The results are contained in Table \ref{tab:f01370}, middle part, and the corresponding CDF is 
displayed in Fig.~\ref{fig:WidthOff01370}. In case of the standard $\chi^2$ fit, the Birge factor comes out equal to about $1.70$.

Regarding the decay modes of the $f_0(1370)$, the situation is perplexing: sixteen modes are listed in Ref.~\cite{pdg2020}, twelve of which are marked as `seen' (one of them being the $\pi \pi$ decay mode). Inspection of the 
results reveals that very little is known about the branching fraction of $f_0(1370) \to \pi \pi$ (denoted as $\Gamma_{1}/\Gamma$ in Ref.~\cite{pdg2020}): just one result exists (in a usable form), yet its relative uncertainty 
is large; this datapoint is labelled in Ref.~\cite{pdg2020} as BUGG96: $0.26(9)$. In an attempt to acquire more data (regarding the interesting - in the context of Ref.~\cite{matsinos2017} - branching fraction), one notices 
the availability of one datapoint for the branching fraction of $f_0(1370) \to K \bar{K}$, denoted as $\Gamma_{11}/\Gamma$ in Ref.~\cite{pdg2020} ($0.35(13)$), as well as of the four results on the ratio $\Gamma_{11}/\Gamma_{1}$ 
($0.08(8)$, $0.91(20)$, $0.12(6)$, and $0.46(15)(11)$). In spite of the large relative uncertainties, one might attempt to fit these six values with two constants, representing the branching fractions of $f_0(1370) \to \pi \pi$ 
and $f_0(1370) \to K \bar{K}$. Unfortunately, the robust fits are very poor (Birge factors $3.02$ and $3.04$, respectively) and suggest that the $\Gamma_{11}/\Gamma_{1}$ results labelled as ABLIKIM05 and ANISOVICH02D be treated 
as outliers. The fitted result for the branching fraction of $f_0(1370) \to \pi \pi$ comes out equal to $30(26)~\%$, whereas the one for the branching fraction of $f_0(1370) \to K \bar{K}$ to $22(24)~\%$; evidently, neither 
result is satisfactory (in terms of accuracy). Although one could opt for the direct use of the BUGG96 $\Gamma_{1}/\Gamma$ result, I would rather refrain from recommending any value for the branching fraction of $f_0(1370) \to \pi \pi$ 
in Tables \ref{tab:ScalarIsoscalar} and \ref{tab:f01370}. I base this decision on my impression regarding the overall reliability of the available data, both in terms of compatibility as well as accuracy, and the very poor 
quality of the robust fits. In view of this outcome, the inclusion of the $t$-channel exchange of an $f_0(1370)$ in the $\pi N$ interaction model of Ref.~\cite{matsinos2017} will be avoided at this time.

In summary, the recommended value of this work for the mass of the $f_0(1370)$ is $1320(12)$ MeV, i.e., considerably more precise than the rough guess at the range between $1200$ and $1500$ MeV, recommended by the PDG. The 
analysis of the available data suggests that the total decay width of this resonance can also be better restricted than the $200-500$ MeV range of values, recommended by the PDG. The decay of the $f_0(1370)$ to $\pi \pi$ 
has been observed, yet the sizeable variability of the available six estimates (which also involved the branching fraction of $f_0(1370) \to K \bar{K}$) can hardly lead to a meaningful determination of the branching fraction 
of $f_0(1370) \to \pi \pi$.

\subsubsection{\label{sec:f01500}$f_0(1500)$}

\begin{figure}
\begin{center}
\includegraphics [width=15.5cm] {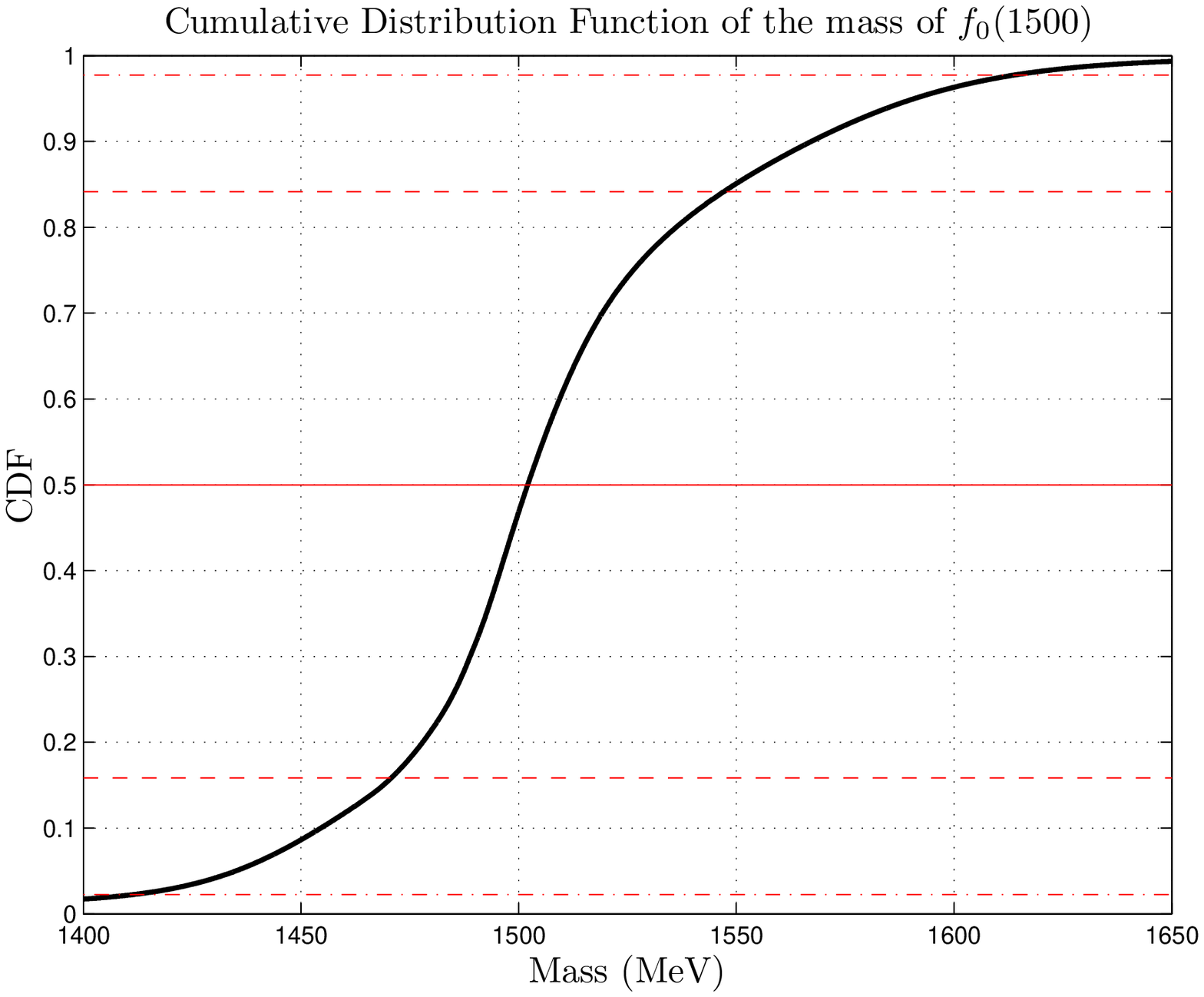}
\caption{\label{fig:MassOff01500}The equivalent of Fig.~\ref{fig:MassOff0500} for the mass of the $f_0(1500)$.}
\vspace{0.35cm}
\end{center}
\end{figure}

The $f_0(1500)$ is the lowest-mass scalar-isoscalar meson for which the PDG reported weighted averages (for the mass and for the total decay width). However, out of the $53$ available results for the mass of this resonance, 
the PDG selected (and based their average of $1506(6)$ MeV on) six; it is perplexing why so many datapoints (including seventeen values, which are more recent than their most recent input result) need take no part in their 
analysis. All datapoints were subjected to the robust optimisation of this work, yielding the results of Table \ref{tab:f01500}, upper part, as well as eight identified outliers, labelled in Ref.~\cite{pdg2020} as DOBBS15 
($1442(9)(4)$ MeV), AAIJ14BR, UMAN06, ANTINORI95 ($1445(5)$ MeV), ABATZIS94, ARMSTRONG89E, ALDE88, and GRAY83. The fit using the standard $\chi^2$ MF on the trimmed dataset of $45$ datapoints yielded the Birge factor of 
about $1.55$. The recommended value of this work for the mass of the $f_0(1500)$ is about $1 \sigma$ (combined uncertainty) below the PDG average. The CDF of the mass of the $f_0(1500)$ is displayed in Fig.~\ref{fig:MassOff01500}.

\begin{table}%[h!]
{\bf \caption{\label{tab:f01500}}}The equivalent of Table \ref{tab:f0500} for the $f_0(1500)$. Regarding the methods featuring the $\chi^2$ MF and the CDF, eight outliers in case of the mass and three in case of the total 
decay width, identified as such by the robust-optimisation methods of Andrews and Tukey, were removed from the original input dataset.
\vspace{0.3cm}
\begin{center}
\begin{tabular}{|c|c|}
\hline
Source/Method & Result\\
\hline
\hline
\multicolumn{2}{|c|}{Mass (MeV)}\\
\hline
PDG \cite{pdg2020} & $1506 \pm 6$\\
Andrews weights & $1498.5^{+4.4}_{-4.2}$\\
Tukey weights & $1498.5^{+4.3}_{-4.2}$\\
Standard $\chi^2$ & $1497.9 \pm 2.8$\\
CDF & $1502.1^{+6.7}_{-4.7}$\\
\hline
\multicolumn{2}{|c|}{Total decay width (MeV)}\\
\hline
PDG \cite{pdg2020} & $112 \pm 9$\\
Andrews weights & $117.2 \pm 4.1$\\
Tukey weights & $117.2 \pm 4.1$\\
Standard $\chi^2$ & $117.7 \pm 3.2$\\
CDF & $121.7^{+8.5}_{-4.7}$\\
\hline
\multicolumn{2}{|c|}{Branching fraction of $f_0(1500) \to \pi \pi$ ($\%$)}\\
\hline
PDG \cite{pdg2020} & $34.5 \pm 2.2$\\
Andrews weights & $28.2^{+5.3}_{-2.8}$\\
Tukey weights & $28.2^{+7.0}_{-2.8}$\\
Standard $\chi^2$ & $28.1^{+2.1}_{-1.9}$\\
\hline
\end{tabular}
\end{center}
\vspace{0.5cm}
\end{table}

\begin{figure}
\begin{center}
\includegraphics [width=15.5cm] {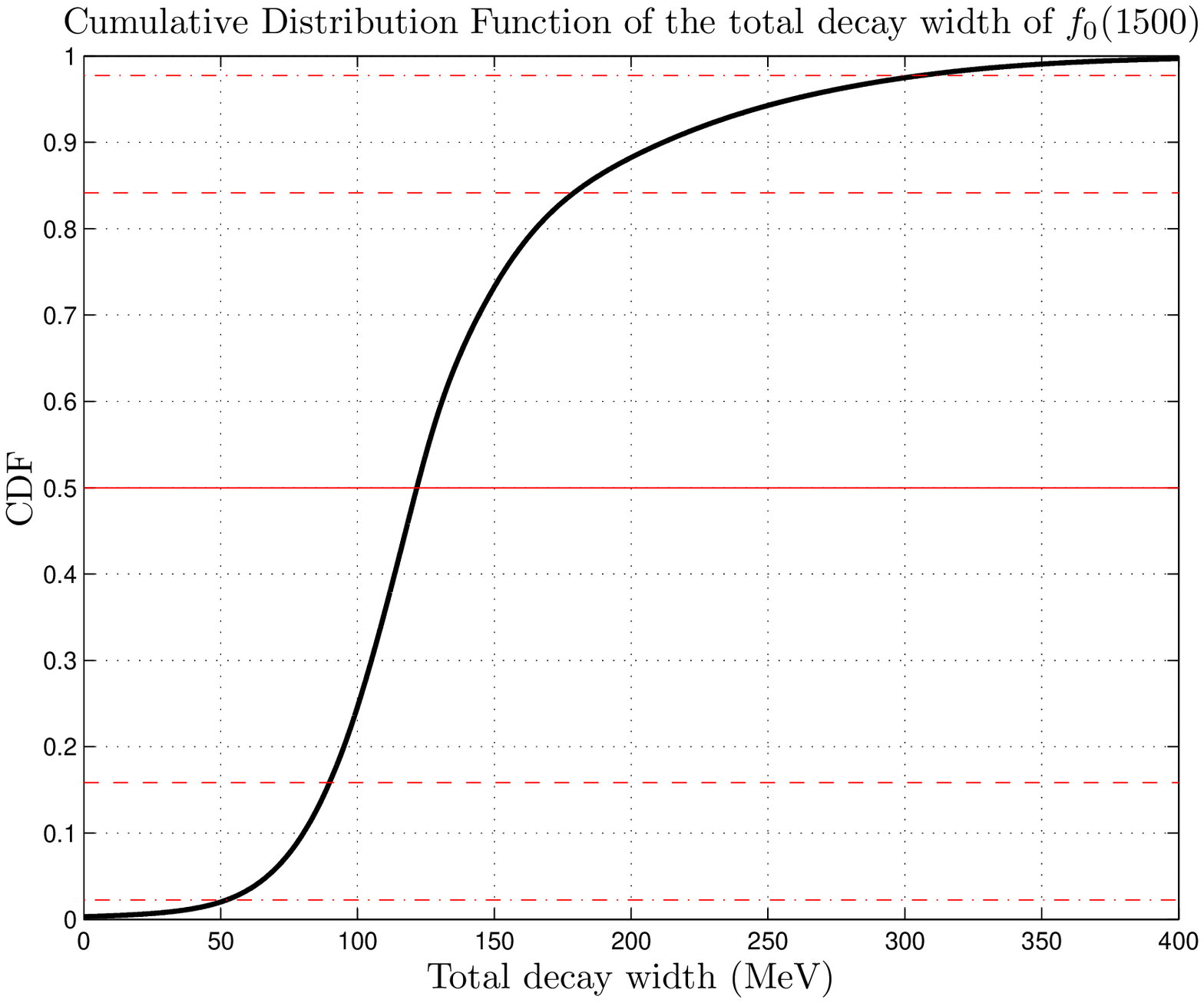}
\caption{\label{fig:WidthOff01500}The equivalent of Fig.~\ref{fig:WidthOff0500} for the total decay width of the $f_0(1500)$.}
\vspace{0.35cm}
\end{center}
\end{figure}

The PDG also selected for analysis six (out of $49$ reported) results on the total decay width of the $f_0(1500)$. However, the two robust methods identified three outliers, labelled in Ref.~\cite{pdg2020} as AUBERT06O, 
ANTINORI95 ($65(10)$ MeV), and ABATZIS94. The standard $\chi^2$ fit to the trimmed dataset yields the acceptable $\chi^2_{\rm min}$ result of about $66.20$ for $45$ DoF, corresponding to $\chi^2_{\rm min}/{\rm NDF} \approx 1.47$ 
and ${\rm p} \approx 2.15 \cdot 10^{-2}$. The CDF of the total decay width of the $f_0(1500)$ is displayed in Fig.~\ref{fig:WidthOff01500}.

I will report on the branching fractions of the $f_0(1500)$ in somewhat more detail. Thirteen `seen' (as well as one `unseen') decay modes of the $f_0(1500)$ are listed in Ref.~\cite{pdg2020}, the dominant ones being to the 
$\pi \pi$, $4 \pi$, $\eta \eta$, $\eta \eta^\prime (958)$, and $K \bar{K}$ final states. The corresponding branching fractions are denoted therein as $\Gamma_{1}/\Gamma$, $\Gamma_{4}/\Gamma$, $\Gamma_{11}/\Gamma$, $\Gamma_{12}/\Gamma$, 
and $\Gamma_{13}/\Gamma$; I will retain their notation. The relevant available data, as they appear in Ref.~\cite{pdg2020}, are as follows.
\begin{itemize}
\item $\Gamma_{1}/\Gamma$. One datapoint: $0.454(104)$ (BUGG96), not included in the fit in Ref.~\cite{pdg2020}.
\item $\Gamma_{13}/\Gamma$. One datapoint: $0.044(21)$ (BUGG96), not included in the fit in Ref.~\cite{pdg2020}.
\item $\Gamma_{4}/\Gamma_{1}$. Four datapoints: $1.37(16)$ (BARBERIS00D), $2.1(6)$ (AMSLER98), $2.1(2)$ (ANISOVICH02D), and $3.4(8)$ (ABELE96). The last two datapoints were not included in the fit in Ref.~\cite{pdg2020}.
\item $\Gamma_{11}/\Gamma_{1}$. Six datapoints: $0.18(3)$ (BARBERIS00E), $0.157(60)$ (AMSLER95D), $0.080(33)$ (AMSLER02), $0.11(3)$ (ANISOVICH02D), $0.078(13)$ (ABELE96C), and $0.230(97)$ (AMSLER95C). The last four datapoints 
were not included in the fit in Ref.~\cite{pdg2020}.
\item $\Gamma_{12}/\Gamma_{1}$. Two datapoints: $0.095(26)$ (BARBERIS00A) and $0.005(3)$ (ANISOVICH02D). The second datapoint was not included in the fit in Ref.~\cite{pdg2020}.
\item $\Gamma_{13}/\Gamma_{1}$. Five datapoints: $0.25(3)$ (BARGIOTTI03), $0.19(7)$ (ABELE98), $0.20(8)$ (ABELE96B), $0.16(5)$ (ANISOVICH02D), and $0.33(3)(7)$ (BARBERIS99D). The last two datapoints were not included in the 
fit in Ref.~\cite{pdg2020}.
\item $\Gamma_{12}/\Gamma_{11}$. Four datapoints: $0.29(10)$ (AMSLER95C), $0.05(3)$ (ANISOVICH02D), $0.84(23)$ (ABELE96C), and $2.7(8)$ (BINON84C). The last three datapoints were not included in the fit in Ref.~\cite{pdg2020}.
\item $\Gamma_{13}/\Gamma_{11}$. Two datapoints: $1.85(41)$ (BARBERIS00E) and $1.5(6)$ (ANISOVICH02D). The second datapoint was not included in the fit in Ref.~\cite{pdg2020}.
\end{itemize}
Therefore, the PDG select ten datapoints as input into their constrained fit featuring four free parameters; and they omit fifteen.

All $25$ datapoints were submitted to the optimisation using Andrews and Tukey weights, and four free parameters, namely the branching fractions $\Gamma_{1}/\Gamma$, $\Gamma_4/\Gamma$, $\Gamma_{11}/\Gamma$, and $\Gamma_{12}/\Gamma$; 
given the constraint $\Gamma_{1}+\Gamma_{4}+\Gamma_{11}+\Gamma_{12}+\Gamma_{13}=\Gamma$, the branching fraction $\Gamma_{13}/\Gamma$ is determined (at any given step of the optimisation) from the free parameters. The two 
robust methods identified four outliers: the $\Gamma_{4}/\Gamma_{1}$ BARBERIS00D result, the $\Gamma_{12}/\Gamma_{1}$ ANISOVICH02D result, and the $\Gamma_{12}/\Gamma_{11}$ AMSLER95C and ANISOVICH02D results; the first and 
the third of these outliers are two (of the ten) datapoints selected for analysis by the PDG. The fitted results for $\Gamma_{1}/\Gamma$ can be found in Table \ref{tab:f01500}, lower part. For the sake of completeness, I also 
give fitted results for the remaining three free parameters in case of the robust-optimisation method using Andrews weights: $\Gamma_{4}/\Gamma = 60.4^{+3.9}_{-7.3}~\%$, $\Gamma_{11}/\Gamma = 2.89^{+0.82}_{-0.62}~\%$, and 
$\Gamma_{12}/\Gamma = 2.68^{+1.02}_{-0.95}~\%$. It follows that the recommended value of this work for $\Gamma_{1}/\Gamma$ is about $1 \sigma$ (combined uncertainty) below the PDG value, whereas its recommended value for 
$\Gamma_{4}/\Gamma$ exceeds the one by the PDG (namely $48.9 \pm 3.3~\%$) by a slightly larger amount. The fitted values of the branching fractions $\Gamma_{11}/\Gamma$ and $\Gamma_{12}/\Gamma$ came out about equal herein; 
the PDG recommend a sizeably higher value in the former case. The average value of the $\Gamma_{13}/\Gamma$ in this work is about $5.86~\%$, not far from the PDG result of $8.5 \pm 1.0~\%$. The Hessian matrix in case of the 
robust-optimisation method using Andrews weights is detailed in Table \ref{tab:Hessianf01500}. Finally, the fit using the standard $\chi^2$ MF in case of the trimmed dataset of $21$ datapoints yields acceptable results: 
$\chi^2_{\rm min} \approx 30.54$ for $17$ DoF, corresponding to the p-value of about $2.27 \cdot 10^{-2}$.

\begin{table}%[h!]
{\bf \caption{\label{tab:Hessianf01500}}}
\vspace{0.3cm}
\begin{center}
\begin{tabular}{|c|cccc|}
\hline
 & $\Gamma_{1}/\Gamma$ & $\Gamma_{4}/\Gamma$ & $\Gamma_{11}/\Gamma$ & $\Gamma_{12}/\Gamma$\\
\hline
$\Gamma_{1}/\Gamma$ & $1.000$ & $-0.858$ & $0.251$ & $0.130$\\
$\Gamma_{4}/\Gamma$ & $-0.858$ & $1.000$ & $-0.574$ & $-0.442$\\
$\Gamma_{11}/\Gamma$ & $0.251$ & $-0.574$ & $1.000$ & $0.346$\\
$\Gamma_{12}/\Gamma$ & $0.130$ & $-0.442$ & $0.346$ & $1.000$\\
\hline
\end{tabular}
\end{center}
\vspace{0.5cm}
\end{table}

In summary, the decision by the PDG to obtain estimates for the mass and for the total decay width of the $f_0(1500)$ from just six results in each case, whereas $53$ datapoints are available in the former case and $49$ in 
the latter, is hard to follow. It is perplexing that seventeen of the omitted results in the former case (and fourteen in the latter) are more recent than the most recent of the data they have decided to consider. The 
recommended values of this work, detailed in Tables \ref{tab:ScalarIsoscalar} and \ref{tab:f01500}, are based on all available data.

Finally, differences were also established in the results of the analysis of the branching fractions of the $f_0(1500)$. The problem with the PDG recommended values is that they rely on ten datapoints, two of which were 
established as outliers in this work. The $\chi^2$ fit to twice as many datapoints as they used yields acceptable results.

\subsubsection{\label{sec:f01710}$f_0(1710)$}

\begin{figure}
\begin{center}
\includegraphics [width=15.5cm] {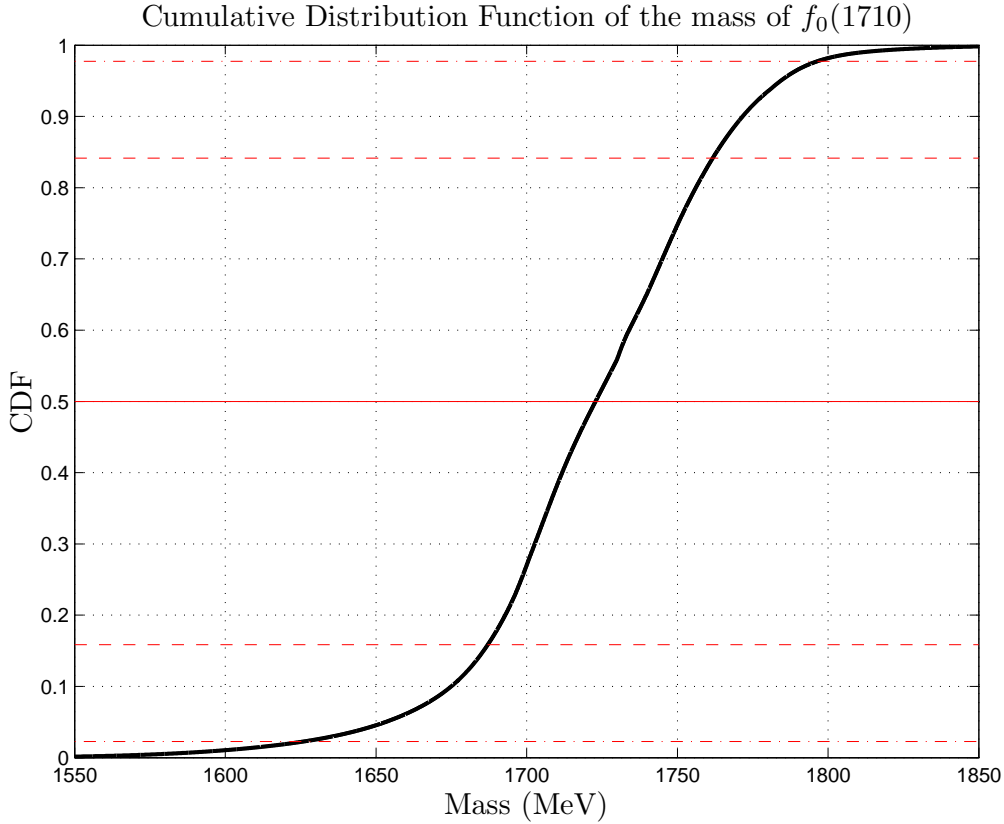}
\caption{\label{fig:MassOff01710}The equivalent of Fig.~\ref{fig:MassOff0500} for the mass of the $f_0(1710)$.}
\vspace{0.35cm}
\end{center}
\end{figure}

Out of the $53$ available estimates for the mass of the $f_0(1710)$, eight were selected for analysis in Ref.~\cite{pdg2020}, yielding the result $1732^{+9}_{-7}$ MeV. On the contrary, all values were submitted to the robust 
optimisation, leading to the identification of seven outliers, labelled in Ref.~\cite{pdg2020} as UMAN06, ANISOVICH99B, BARKOV98, BUGG95 ($1620(16)$ MeV), FALVARD88 (both estimates), and ALDE86C. These seven results were 
removed from the original input database and the remaining data were analysed with the two non-robust methods of this work, yielding the results of Table \ref{tab:f01710}, upper part. In case of the standard $\chi^2$ fit, 
the Birge factor comes out equal to about $1.88$. The difference between the recommended values of this work and of Ref.~\cite{pdg2020} amounts to about $1 \sigma$ (combined uncertainty); this work suggests a lower mass 
for the $f_0(1710)$. The CDF of the mass of the $f_0(1710)$ is displayed in Fig.~\ref{fig:MassOff01710}.

\begin{table}%[h!]
{\bf \caption{\label{tab:f01710}}}The equivalent of Table \ref{tab:f0500} for the $f_0(1710)$. Regarding the methods featuring the $\chi^2$ MF and the CDF, seven outliers in case of the mass and eight in case of the total 
decay width, identified as such by the robust-optimisation methods of Andrews and Tukey, were removed from the original input dataset. Included in this table are the PDG weighted averages; their estimates can be found in 
Table \ref{tab:ScalarIsoscalar} (see also caption of that table).
\vspace{0.3cm}
\begin{center}
\begin{tabular}{|c|c|}
\hline
Source/Method & Result\\
\hline
\hline
\multicolumn{2}{|c|}{Mass (MeV)}\\
\hline
PDG \cite{pdg2020} & $1732^{+9}_{-7}$\\
Andrews weights & $1715.1^{+10.5}_{-9.2}$\\
Tukey weights & $1715.3^{+10.9}_{-8.9}$\\
Standard $\chi^2$ & $1720.4 \pm 3.9$\\
CDF & $1722.8^{+5.7}_{-5.3}$\\
\hline
\multicolumn{2}{|c|}{Total decay width (MeV)}\\
\hline
PDG \cite{pdg2020} & $147^{+12}_{-10}$\\
Andrews weights & $144 \pm 13$\\
Tukey weights & $144^{+12}_{-13}$\\
Standard $\chi^2$ & $150.0 \pm 6.4$\\
CDF & $147.0^{+12.8}_{-8.3}$\\
\hline
\multicolumn{2}{|c|}{Branching fraction of $f_0(1710) \to \pi \pi$ ($\%$)}\\
\hline
PDG \cite{pdg2020} & $-$\\
Andrews weights & $8.9^{+3.2}_{-3.0}$\\
Tukey weights & $8.9^{+3.2}_{-3.0}$\\
Standard $\chi^2$ & $8.9 \pm 2.3$\\
\hline
\end{tabular}
\end{center}
\vspace{0.5cm}
\end{table}

\begin{figure}
\begin{center}
\includegraphics [width=15.5cm] {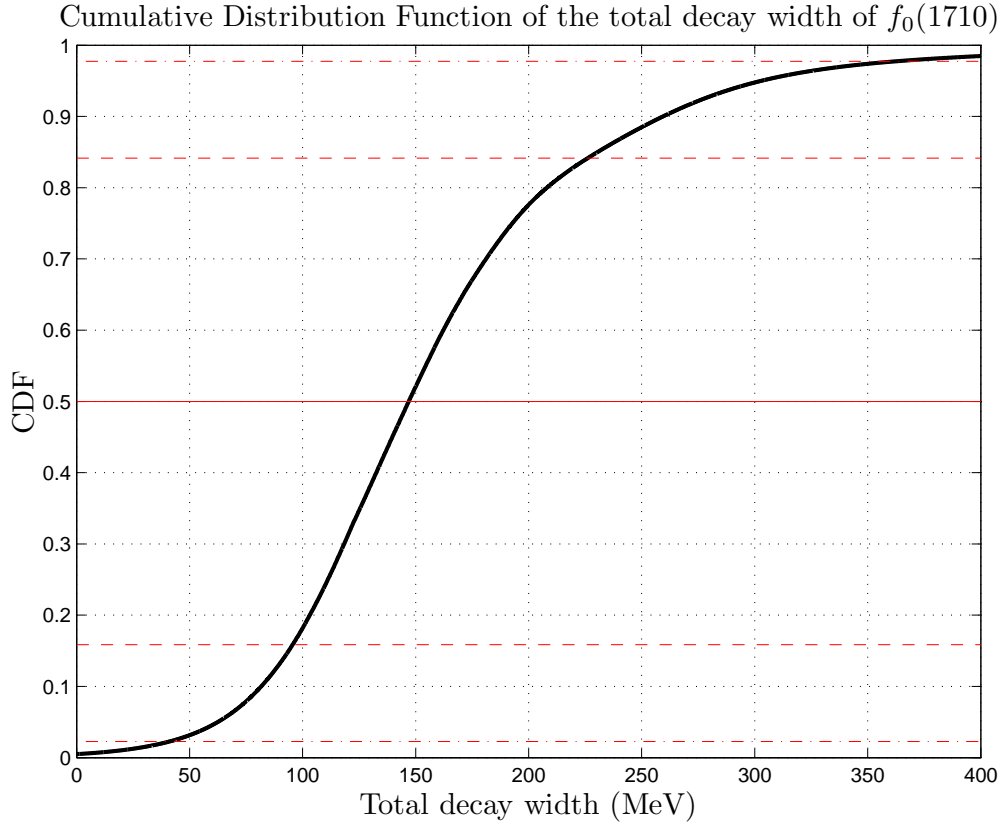}
\caption{\label{fig:WidthOff01710}The equivalent of Fig.~\ref{fig:WidthOff0500} for the total decay width of the $f_0(1710)$.}
\vspace{0.35cm}
\end{center}
\end{figure}

The PDG selected for analysis also eight (out of $46$ reported) results on the total decay width of the $f_0(1710)$. The two robust methods identified eight outliers, labelled in Ref.~\cite{pdg2020} as ABLIKIM05, ANISOVICH03 
($320^{+50}_{-20}$ MeV), BARKOV98, BALOSHIN95, ARMSTRONG93C, BOLONKIN88 ($30(20)$ MeV), FALVARD88 ($184(6)$ MeV), and ETKIN82B. The standard $\chi^2$ fit to the trimmed dataset of $38$ datapoints yielded the Birge factor 
of about $1.44$. The CDF of the total decay width of the $f_0(1710)$ is displayed in Fig.~\ref{fig:WidthOff01710}.

It must be mentioned that the PDG provide two results both for the mass as well as for the total decay width of the $f_0(1710)$: one corresponding to their weighted average (result given in Table \ref{tab:f01710}), the other 
obtained ``not from a direct measurement, but evaluated from measurements of related quantities'' \cite{pdg2020}. Their latter results point to lower values both for the mass as well as for the total decay width of the 
$f_0(1710)$, see Table \ref{tab:ScalarIsoscalar}.

Five decay modes of the $f_0(1710)$ are mentioned in Ref.~\cite{pdg2020}: to the $K \bar{K}$, $\eta \eta$, $\pi \pi$, $\gamma \gamma$, and $\omega \omega$ final states, represented in Ref.~\cite{pdg2020} by the branching 
fractions $\Gamma_{1, \dots , 5}/\Gamma$, respectively. Although the decay mode $f_0(1710) \to \omega \omega$ is marked in Ref.~\cite{pdg2020} as `seen', no data can be found for the relevant branching fraction $\Gamma_{5}/\Gamma$, 
direct or indirect (i.e., involving ratios to other branching fractions). From the only available value for the product $\Gamma_{1} \Gamma_{4} / \Gamma$ ($12^{+3 \,\, +227}_{-2 \,\, -8}$ eV) as well as from the upper limit 
of the product $\Gamma_{3} \Gamma_{4} / \Gamma$ ($< 0.82$ keV at $95~\%$ confidence), it follows that the decay mode to $\gamma \gamma$ can safely be omitted. As a consequence, the useful data may be summarised as follows.
\begin{itemize}
\item $\Gamma_{1}/\Gamma$. Two datapoints: $0.36(12)$ (ALBALADEJO08) and $0.38^{+0.09}_{-0.19}$ (LONGACRE86).
\item $\Gamma_{2}/\Gamma$. Two datapoints: $0.22(12)$ (ALBALADEJO08) and $0.18^{+0.03}_{-0.13}$ (LONGACRE86).
\item $\Gamma_{3}/\Gamma$. One datapoint: $0.039^{+0.002}_{-0.024}$ (LONGACRE86).
\item $\Gamma_{2}/\Gamma_{1}$. Two datapoints: $0.48(15)$ (BARBERIS00E) and $0.46^{+0.70}_{-0.38}$ (ANISOVICH02D).
\item $\Gamma_{3}/\Gamma_{1}$. Five datapoints: $0.64(27)(18)$ (LEES18A), $0.41^{+0.11}_{-0.17}$ (ABLIKIM06V), $0.200(24)(36)$ BARBERIS99D, $0.39(14)$ (ARMSTRONG91), and $0.32(14)$ (ALBALADEJO08). (Due to its enormous 
uncertainties, the ANISOVICH02D estimate will be omitted.)
\end{itemize}
The description of these twelve datapoints with three free parameters, namely $\Gamma_{1, 2, 3}/\Gamma$, was first investigated with the robust fits. Both statistical methods identified the $\Gamma_{3}/\Gamma$ datapoint as 
outlier. The exclusion of this datapoint led to the $\chi^2_{\rm min}$ value of $4.97$ for $8$ DoF; the smallness of this result may be traced to the large uncertainties of the input data. The fitted results for the branching 
fraction $\Gamma_{3}/\Gamma$ show no dependence on the method of the fit, see Table \ref{tab:f01710}, lower part. For the sake of completeness, I will also give the fitted results for the parameters $\Gamma_{1}/\Gamma$ and 
$\Gamma_{2}/\Gamma$ from the robust-optimisation method using Andrews weights: $\Gamma_{1}/\Gamma=37.1^{+8.4}_{-10.6}~\%$ and $\Gamma_{2}/\Gamma=18.1^{+3.4}_{-6.2}~\%$. These three estimates leave room for the branching 
fraction $\Gamma_{5}/\Gamma$ (about $35.8~\%$, with a sizeable relative uncertainty). Two remarks are due. First, given the expected magnitude of the branching fraction $\Gamma_{5}/\Gamma$, it is surprising that no information 
appears to be available on the decay mode of the $f_0(1710)$ to $\omega \omega$. Second, the interesting branching fraction (in the context of Ref.~\cite{matsinos2017}) should be close to $10~\%$.

In summary, the two recommended values for the mass of the $f_0(1710)$ - the one from this work and the other from the most recent PDG compilation - are about $1 \sigma$ apart; this work suggests a lower value. The two 
recommended values for the total decay width of this resonance agree, whereas the interesting branching fraction (in the context of Ref.~\cite{matsinos2017}), not given in Ref.~\cite{pdg2020}, should be close to about $10~\%$.

\subsubsection{\label{sec:ConclusionsScalarIsoscalar}Concluding remarks on the scalar-isoscalar mesons}

One last step was taken towards the consolidation of the results of this work: all normalised residuals from the $\chi^2$ fits~\footnote{The application of distance-dependent weights obfuscates the interpretation of the 
corresponding results in case of the two robust-optimisation methods.} were submitted to the Shapiro-Wilk test, an established method in testing whether a set of values have been sampled from a Gaussian distribution. 
Originally introduced in 1965 \cite{Shapiro1965} for small samples (containing up to $50$ datapoints), the Shapiro-Wilk test was extended in a series of papers (mostly by Royston), to be applicable to sets containing up to 
$5\,000$ elements by the mid 1990s \cite{Royston1995}. Its results are easy to interpret: the output of the application of the test is one value, the so-called $W$-statistic~\footnote{One of the standard ways in Statistics 
to demonstrate that one physical quantity follows a given distribution is to investigate the relation between the quantiles of the distribution of that quantity (e.g., the values of the physical quantity corresponding to the 
$10, 20, \dots , 90~\%$ levels of its CDF) and those of the given distribution. If the linearity between the two sets of quantiles cannot be refuted, then the physical quantity in question is accepted as following the given 
distribution. As Royston comments in Ref.~\cite{Royston1995}, the $W$-statistic is an approximate measure of the linearity between the quantiles of the distribution which is being under test and those of the normal 
distribution.}, which is translated (via well-established transformations) into the probability that the input dataset is normally distributed (p-value for accepting the null hypothesis). The method may be implemented in 
such a way as to also admit (as input) the (user-defined) maximal number of potential outliers (i.e., the maximal number of elements of the original set which could be exempt from the test (no contribution to the 
$W$-statistic). In this work, the results correspond to \emph{no outliers}, which implies that each test would have failed if it returned a p-value smaller than ${\rm p}_{\rm min}$, and that no further attempt would have 
been made in order to improve the outcome of the test by removing datapoints from the original distribution of the normalised residuals. The values of the $W$-statistic and the p-values from the Shapiro-Wilk test in case of 
the masses and of the total decay widths of the five scalar-isoscalar mesons of this section can be found in Table \ref{tab:SWTestScalarIsoscalar}. In short, there is no indication that any of the examined ten distributions 
departs from the Gaussian distribution at a statistically significant level; the minimal p-value from these tests exceeds $0.20$.

\begin{table}%[h!]
{\bf \caption{\label{tab:SWTestScalarIsoscalar}}}The results of the application of the Shapiro-Wilk test \cite{Shapiro1965,Royston1995} to the normalised residuals of the $\chi^2$ fits to the trimmed datasets containing 
the estimates for the mass and for the total decay width of each scalar-isoscalar meson.
\vspace{0.3cm}
\begin{center}
\begin{tabular}{|c|c|c|}
\hline
Case & $W$-statistic & p-value\\
\hline
\hline
Mass of the $f_0(500)$ & $0.976$ & $0.697$\\
Total decay width of the $f_0(500)$ & $0.904$ & $0.313$\\
\hline
Mass of the $f_0(980)$ & $0.982$ & $0.572$\\
Total decay width of the $f_0(980)$ & $0.966$ & $0.233$\\
\hline
Mass of the $f_0(1370)$ & $0.976$ & $0.617$\\
Total decay width of the $f_0(1370)$ & $0.953$ & $0.381$\\
\hline
Mass of the $f_0(1500)$ & $0.986$ & $0.870$\\
Total decay width of the $f_0(1500)$ & $0.976$ & $0.467$\\
\hline
Mass of the $f_0(1710)$ & $0.970$ & $0.278$\\
Total decay width of the $f_0(1710)$ & $0.971$ & $0.420$\\
\hline
\end{tabular}
\end{center}
\vspace{0.5cm}
\end{table}

The overall picture would have been satisfactory, save for one hitch, namely that only two of the p-values from the $\chi^2$ fits to the same data exceed the ${\rm p}_{\rm min}$ threshold of this work: the lowest p-value 
of $6.01 \cdot 10^{-15}$ was obtained in the fit to the mass values of the $f_0(980)$. This evident mismatch brings up the plausible question: How can the p-values from the $\chi^2$ fits attest to the poor quality of the 
data description in eight cases, while the results of the application of the Shapiro-Wilk test of Table \ref{tab:SWTestScalarIsoscalar} are so satisfactory? The answer to this question is quite simple, and may be found at 
the end of Section \ref{sec:MethodsII}. The application of the two robust-optimisation methods leads to the removal of the distant outliers. The trimmed datasets still contain datapoints which contribute significantly to 
the standard $\chi^2$ MF, yet such datapoints can only be, according to the results of the Shapiro-Wilk test, symmetrically distributed about the $\avg{y}$ fitted results. As mentioned in Section \ref{sec:MethodsII}, one 
could aim at removing additional datapoints from the original input datasets, yet such a move is unnecessary: provided that the `surviving' outliers are symmetrically distributed about the $\avg{y}$ fitted results, there is 
no danger that the minimisation algorithms might drift towards one of the tails of the distribution (of the input values). Naturally, one must correct (enlarge) the fitted uncertainties $\delta \avg{y}$, so that they also 
take account of the fit quality; this assignment is left to the Birge factor.

To summarise, the distribution of the normalised residuals in eight (out of the ten) cases in this chapter is surely not the normal distribution $N(\mu=0,\sigma^2=1)$ (as one would ideally expect on the basis of pure 
statistical fluctuations): however, the Shapiro-Wilk test demonstrates that it is a Gaussian, i.e., $N(\mu=0,\sigma^2>1)$. Via the redefinition of the weights $w_i$ of the input datapoints, the application of the Birge 
factor has the practical effect of transforming each such Gaussian into the normal distribution.

\subsection{\label{sec:VectorIsovector}Vector-isovector mesons}

This section relates to the properties of the vector-isovector mesons. The summary of the recommended values can be found in Table \ref{tab:VectorIsovector}. Details about the data analysis can be found in Sections 
\ref{sec:rho770}-\ref{sec:rho1900}. Given that the two-pion decay of two of these resonances has not been established, and that no data is available in a third case, emphasis in this section will be placed on the two 
resonances which are useful in the context of the $\pi N$ interaction model of Ref.~\cite{matsinos2017}; the remaining three cases will be treated rather epigrammatically.

The decays of all neutral vector-isovector mesons into two neutral pions, $\rho^0(\dots) \to \pi^0 \pi^0$, are forbidden in the framework of the SU(2) isospin symmetry: the Clebsch-Gordan coefficient 
$\bra{I_1, I_2 ; I_{1z}, I_{2z}}\ket{I ; I_z}$, where $I_1=I_2=1$ and $I_{1z}=I_{2z}=0$ (two $\pi^0$'s in the final state), and $I=1$ and $I_z=0$ ($\rho^0$ in the initial state) vanishes.

\begin{table}%[h!]
{\bf \caption{\label{tab:VectorIsovector}}}Summary of the recommended values for the physical properties of the vector-isovector mesons below $2$ GeV. The masses and the total decay widths are expressed in MeV, the branching 
fractions in percent. The PDG notation for the method yielding their results is as follows. AVERAGE: from a weighted average of selected data. ESTIMATE: based on the observed range of the data; not from a formal statistical 
procedure. FIT: from a constrained or overdetermined multiparameter fit of selected data.
\vspace{0.3cm}
\begin{center}
\begin{tabular}{|c|c|c|}
\hline
Physical quantity & PDG \cite{pdg2020} & This work\\
\hline
\hline
\multicolumn{3}{|c|}{$\rho^\pm(770)$}\\
\hline
Mass & $775.11 \pm 0.34$ (AVERAGE) & $775.12 \pm 0.31$\\
Total decay width & $149.1 \pm 0.8$ (AVERAGE, FIT) & $148.0^{+1.6}_{-1.7}$\\
Branching fraction of $\rho^\pm(770) \to \pi^\pm \pi^0$ & $\approx 100$ & $\approx 99$\\
\hline
\multicolumn{3}{|c|}{$\rho^0(770)$}\\
\hline
Mass & $775.26 \pm 0.25$ (AVERAGE) & $775.38^{+0.48}_{-0.46}$\\
Total decay width & $147.8 \pm 0.9$ (AVERAGE) & $145.30^{+0.73}_{-0.70}$\\
Branching fraction of $\rho^0(770) \to \pi^+ \pi^-$ & $\approx 100$ & $\approx 99$\\
\hline
\multicolumn{3}{|c|}{$\rho(1450)$}\\
\hline
Mass & $1465 \pm 25$ (ESTIMATE) & $1421^{+16}_{-19}$\\
Total decay width & $400 \pm 60$ (ESTIMATE) & $399^{+42}_{-44}$\\
Branching fraction of $\rho(1450) \to \pi \pi$ & $-$ & $-$\\
\hline
\multicolumn{3}{|c|}{$\rho(1570)$}\\
\hline
Mass & $1570 \pm 36 \pm 62$ & $-$\\
Total decay width & $144 \pm 75 \pm 43$ & $-$\\
Branching fraction of $\rho(1570) \to \pi \pi$ & $-$ & $-$\\
\hline
\multicolumn{3}{|c|}{$\rho(1700)$}\\
\hline
Mass & $1720 \pm 20$ (ESTIMATE) & $1723^{+29}_{-28}$\\
Total decay width & $250 \pm 100$ (ESTIMATE) & $271 \pm 15$\\
Branching fraction of $\rho(1700) \to \pi \pi$ & $-$ & $27.4^{+2.6}_{-2.7}$\\
\hline
\multicolumn{3}{|c|}{$\rho(1900)$}\\
\hline
Mass & $-$ & $1880 \pm 22$\\
Total decay width & $-$ & $151^{+73}_{-75}$\\
Branching fraction of $\rho(1900) \to \pi \pi$ & $-$ & $-$\\
\hline
\end{tabular}
\end{center}
\vspace{0.5cm}
\end{table}

\subsubsection{\label{sec:rho770}$\rho(770)$}

Regarding the vector-isovector mesons, this is the most important of the $t$-channel exchanges in the $\pi N$ interaction model of Ref.~\cite{matsinos2017}.

The hurdles in the determination of the mass of the $\rho(770)$ are addressed in the introduction of the relevant chapter in Ref.~\cite{pdg2020}, which has been authored by S.~Eidelman and G.~Venanzoni. The authors recommend 
the use of the results from the $e^+ e^-$ annihilation and from the $\tau$-lepton decay for ``the cleanest determination of the $\rho(770)$ mass and width.'' Their recommendation will be followed in this analysis, yet all 
listed values under the `$\rho(770)$-mass/width' Sections `Neutral only, $e^+ e^-$' and `Charged only, $\tau$ decays and $e^+ e^-$' will be imported for analysis. Those of us who were involved in Particle-Physics research in 
the 1980s and 1990s surely recall the lower $\rho(770)$-mass values which were in use at those times, about $1~\%$ below most of the results in the post-LEP era, see also Ref.~\cite{Bertsch1990}.

\begin{figure}
\begin{center}
\includegraphics [width=15.5cm] {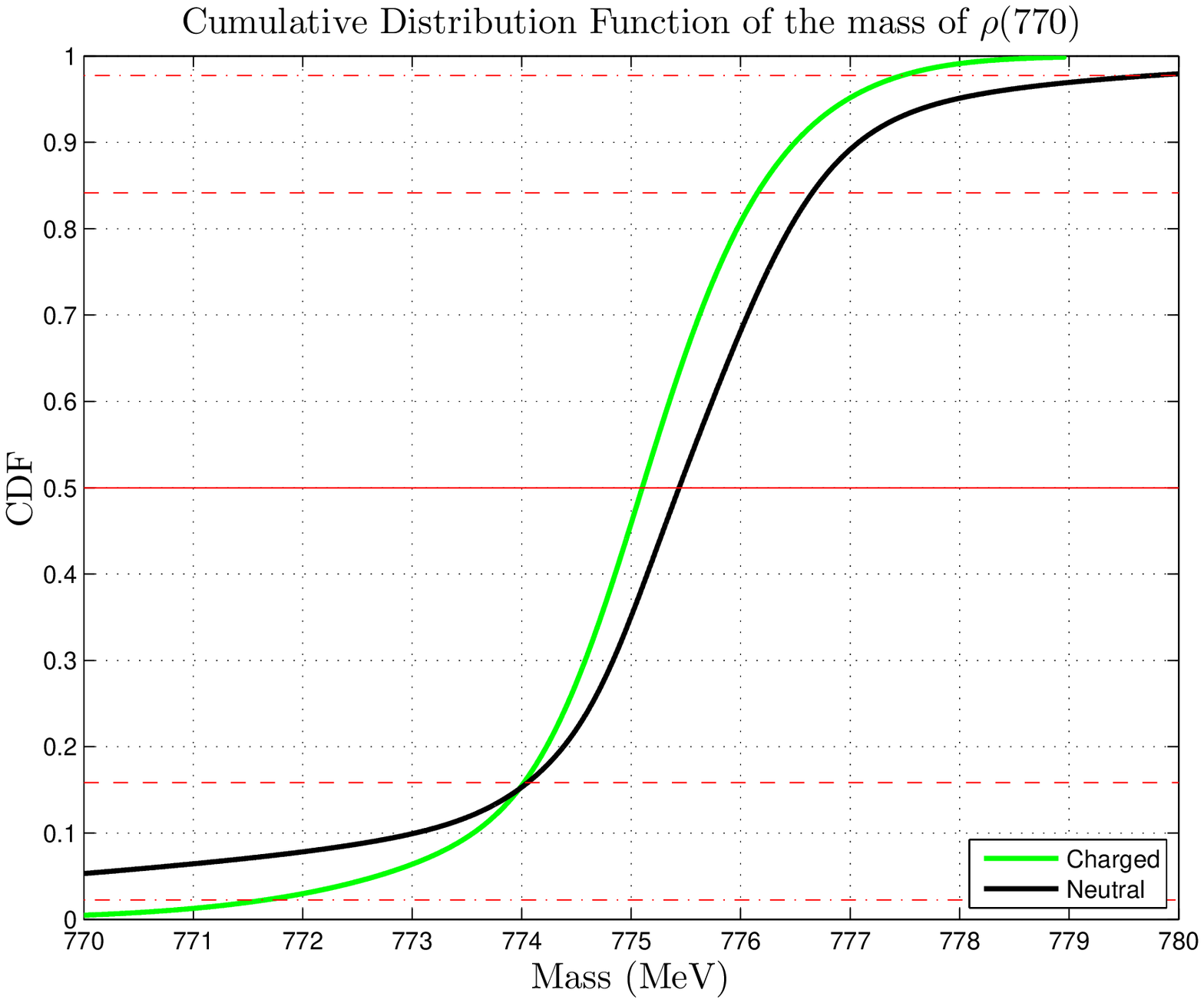}
\caption{\label{fig:MassOfrho770}The CDFs of the mass of the $\rho^\pm(770)$ and of the $\rho^0(770)$, obtained from averaging nine and twelve Gaussian distributions, respectively. One outlier in the former case and five in 
the latter were removed from the original input dataset on the basis of the results of the robust-optimisation methods of Andrews and Tukey. The horizontal solid straight line marks the $50~\%$ level; the two horizontal dashed 
straight lines delineate the $68.27~\%$ CI, the equivalent of $1 \sigma$ limits in the normal distribution; the two horizontal dashed-dotted straight lines delineate the $95.45~\%$ CI, the equivalent of $2 \sigma$ limits in 
the normal distribution.}
\vspace{0.35cm}
\end{center}
\end{figure}

Regarding the mass of the $\rho^\pm(770)$, ten values are listed in Ref.~\cite{pdg2020}, yet only four found their way to the PDG input database, which yielded their recommended result ($775.11(34)$ MeV). The application of 
the two robust methods to the entire set of available data resulted in the fitted values and uncertainties of Table \ref{tab:rho770}, upper part; the agreement with the PDG average is good. The two robust methods identified 
just one outlier, labelled in Ref.~\cite{pdg2020} as BARTOS17A; this datapoint had not been used in the PDG average. After its removal, the dataset was submitted to the optimisation using the standard $\chi^2$ MF, as well as 
to the analysis featuring the determination of the properties of the resonances from the CDF. In the former case, the $\chi^2_{\rm min}$ value came out equal to about $3.92$ for $8$ DoF; one may only speculate about the 
smallness of this $\chi^2_{\rm min}$ result. The CDF of the mass of the $\rho^\pm(770)$ is displayed in Fig.~\ref{fig:MassOfrho770}.

Regarding the mass of the $\rho^0(770)$, seventeen values are listed in Ref.~\cite{pdg2020}, yet only seven were involved in their average of $775.26(25)$ MeV. The application of the two robust methods resulted in the fitted 
values and uncertainties of Table \ref{tab:rho770}, upper part, which are in agreement with the PDG average. The two robust methods identified five outliers, labelled in Ref.~\cite{pdg2020} as BARTOS17 \& 17A, O'CONNELL97, 
BERNICHA94, and GESHKENBEIN89; none of these datapoints had been used in the evaluation of the PDG result. After the removal of the outliers, the dataset was submitted to the optimisation using the standard $\chi^2$ MF and to 
the analysis featuring the determination of the properties of the resonances from the CDF. In the former case, the $\chi^2_{\rm min}$ value came out equal to about $5.80$ for $11$ DoF, another `too good to be true' result. 
The CDF of the mass of the $\rho^0(770)$ is also displayed in Fig.~\ref{fig:MassOfrho770}.

\begin{table}%[h!]
{\bf \caption{\label{tab:rho770}}}Physical properties of the $\rho(770)$. The statistical methods using Andrews and Tukey weights are robust and admit the entire set of available results. Regarding the methods featuring the 
$\chi^2$ MF and the CDF in case of the mass, one outlier from the original $\rho^\pm$ and five from the original $\rho^0$ datasets, identified as such by the two aforementioned robust methods, were removed; in case of the 
total decay width, the corresponding numbers are one and two outliers, respectively.
\vspace{0.3cm}
\begin{center}
\begin{tabular}{|c|c|c|}
\hline
Source/Method & Result for $\rho^\pm$ & Result for $\rho^0$\\
\hline
\hline
\multicolumn{3}{|c|}{Mass (MeV)}\\
\hline
PDG \cite{pdg2020} & $775.11 \pm 0.34$ & $775.26 \pm 0.25$\\
Andrews weights & $775.12 \pm 0.31$ & $775.38^{+0.48}_{-0.46}$\\
Tukey weights & $775.12 \pm 0.31$ & $775.38^{+0.48}_{-0.46}$\\
Standard $\chi^2$ & $775.12 \pm 0.25$ & $775.36 \pm 0.21$\\
CDF & $775.10^{+0.35}_{-0.36}$ & $775.44^{+0.35}_{-0.40}$\\
\hline
\multicolumn{3}{|c|}{Total decay width (MeV)}\\
\hline
PDG \cite{pdg2020} & $149.1 \pm 0.8$ & $147.8 \pm 0.9$\\
Andrews weights & $148.0^{+1.6}_{-1.7}$ & $145.30^{+0.73}_{-0.70}$\\
Tukey weights & $148.0^{+1.6}_{-1.7}$ & $145.30^{+0.73}_{-0.70}$\\
Standard $\chi^2$ & $147.40 \pm 0.98$ & $145.33 \pm 0.45$\\
CDF & $148.1^{+1.1}_{-1.5}$ & $146.03^{+1.28}_{-0.66}$\\
\hline
\multicolumn{3}{|c|}{Branching fraction of $\rho(770) \to \pi \pi$ ($\%$)}\\
\hline
PDG \cite{pdg2020} & $\approx 100$ & $\approx 100$\\
This work & $\approx 99$ & $\approx 99$\\
\hline
\end{tabular}
\end{center}
\vspace{0.5cm}
\end{table}

\begin{figure}
\begin{center}
\includegraphics [width=15.5cm] {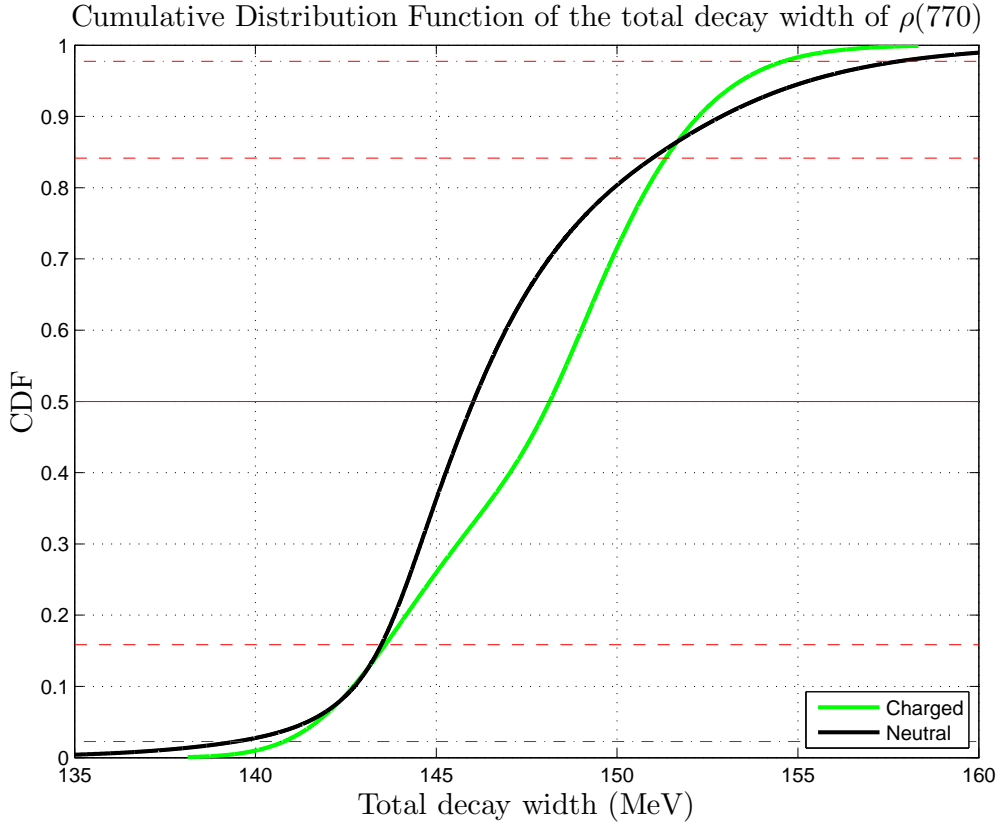}
\caption{\label{fig:WidthOfrho770}The CDFs of the total decay width of the $\rho^\pm(770)$ and of the $\rho^0(770)$, obtained from averaging nine and fifteen Gaussian distributions, respectively. The horizontal straight 
lines represent the same levels as in Fig.~\ref{fig:MassOfrho770}.}
\vspace{0.35cm}
\end{center}
\end{figure}

Regarding the total decay width of the $\rho^\pm(770)$ four (out of the ten available) datapoints were used in Ref.~\cite{pdg2020}. The application of the two robust methods resulted in the identification of one outlier, 
labelled in Ref.~\cite{pdg2020} as BARTOS17A. For the $\rho^0(770)$, used by the PDG were seven (out of the seventeen available) datapoints. However, the robust fits to the available data suggest two outliers~\footnote{The 
first of these datapoints (LEES12G), which - owing to its accuracy - carries a large weight in the optimisation, had been used in the evaluation of the PDG result of Table \ref{tab:rho770}; in fact, the indication that 
something is probably amiss about the LEES12G result may be found in the plot of the PDF of the total decay width of the $\rho^0(770)$ shown in Ref.~\cite{pdg2020}; that datapoint contributes the lion's share (about $7.4$) 
to their $\chi^2_{\rm min} \approx 19.6$.}, namely the datapoints labelled as LEES12G and GESHKENBEIN89. The two CDFs are displayed in Fig.~\ref{fig:WidthOfrho770}. One notices that the CDF, corresponding to the total decay 
width of the $\rho^\pm(770)$, appears to depart from the expectation for a normal CDF. Visual inspection of the corresponding PDF (not shown) suggests that the distribution of the total decay width of the $\rho^\pm(770)$ 
actually comprises two overlapping, yet displaced (by about $5$ MeV) with respect to one another, distributions. On the other hand, the PDF of the total decay width of the $\rho^0$ of this work does not resemble the 
distribution shown in Ref.~\cite{pdg2020}: it is unimodal (the LEES12G datapoint is largely to blame for the bimodal distribution displayed in Ref.~\cite{pdg2020}), but right-skewed (yielding a positive coefficient of skewness).

Inspection of the available information on the decay modes of the $\rho(770)$ reveals that the two-pion decay mode is dominant. In case of the $\rho^\pm(770)$, one branching fraction (to $\pi^\pm \gamma$) is properly 
known (in the sense of the availability of estimates for an average and for a relevant meaningful uncertainty); only upper limits are known for the branching fractions to the $\pi^\pm \eta$ and $\pi^\pm \pi^+ \pi^- \pi^0$ 
final states. It thus appears that a lower limit for the branching fraction of the $\rho^\pm(770)$ decay to $\pi^\pm \pi^0$ is equal to $99.15~\%$ at $84.13~\%$ confidence. In case of the $\rho^0(770)$, the sum of all 
branching fractions, other than to two pions ($\pi^+ \pi^-$), slightly exceeds the one-percent level - $1.09(16)~\%$ - and is mostly due to the decay to $\pi^+ \pi^- \gamma$. Given the overall status, in particular regarding 
the total decay width of the $\rho^\pm(770)$, it might be an idea to accept the PDG recommendation and fix the two-pion branching fraction to $100~\%$ regardless of the $\rho(770)$ electrical charge. Although it might appear 
pedantic and fastidious as a suggestion, the alternative would be to use about $99~\%$ in both cases, thus leaving some room for all other branching fractions.

Henceforth, the estimates of Section \ref{sec:VectorIsovector} will make no distinction among the members of each isospin triplet.

\subsubsection{\label{sec:rho1450}$\rho(1450)$}

For the mass of this resonance, the results are shown in Table \ref{tab:rho1450}, upper part.
\begin{itemize}
\item The application of the two robust methods to the original input dataset of $36$ datapoints yielded eleven outliers, labelled in Ref.~\cite{pdg2020} as ACHASOV18 and AKHMETSHIN01B (from the $\eta \rho^0$ mode), ACHASOV16D, 
AKHMETSHIN03B, and EDWARDS00A (from the $\omega \pi$ mode), ARMSTRONG89E (from the $4 \pi$ mode), BARTOS17, LEES12G, SCHAEL05C, and KURDADZE83 (from the $\pi \pi$ mode), and AAIJ16N (from the $K \bar{K}$ mode).
\item The trimmed dataset was submitted to the optimisation using the standard $\chi^2$ MF and resulted in $\chi^2_{\rm min} \approx 126.35$ for $24$ DoF, i.e., in a Birge factor of about $2.29$.
\item The results from the application of the four statistical methods of this work agree well, and are short of the PDG estimate by about $1.5 \sigma$ (combined uncertainty). The relevant uncertainty in the recommended 
value of this work (result with Andrews weights) is about $30~\%$ smaller than the one in the PDG estimate.
\end{itemize}

\begin{table}%[h!]
{\bf \caption{\label{tab:rho1450}}}The equivalent of Table \ref{tab:rho770} for the $\rho(1450)$. Regarding the methods featuring the $\chi^2$ MF and the CDF, eleven outliers in case of the mass and nine in case of the total 
decay width, identified as such by the robust-optimisation methods of Andrews and Tukey, were removed from the original input dataset.
\vspace{0.3cm}
\begin{center}
\begin{tabular}{|c|c|}
\hline
Source/Method & Result\\
\hline
\hline
\multicolumn{2}{|c|}{Mass (MeV)}\\
\hline
PDG \cite{pdg2020} & $1465 \pm 25$\\
Andrews weights & $1421^{+16}_{-19}$\\
Tukey weights & $1421^{+17}_{-19}$\\
Standard $\chi^2$ & $1419.9 \pm 8.3$\\
CDF & $1409^{+13}_{-18}$\\
\hline
\multicolumn{2}{|c|}{Total decay width (MeV)}\\
\hline
PDG \cite{pdg2020} & $400 \pm 60$\\
Andrews weights & $399^{+42}_{-44}$\\
Tukey weights & $396^{+44}_{-43}$\\
Standard $\chi^2$ & $370 \pm 14$\\
CDF & $390^{+18}_{-24}$\\
\hline
\multicolumn{2}{|c|}{Branching fraction of $\rho(1450) \to \pi \pi$ ($\%$)}\\
\hline
PDG \cite{pdg2020} & $-$\\
This work & $-$\\
\hline
\end{tabular}
\end{center}
\vspace{0.5cm}
\end{table}

For the total decay width of this resonance, the results are shown in Table \ref{tab:rho1450}, middle part.
\begin{itemize}
\item The application of the two robust methods to the original input dataset of $33$ datapoints yielded nine outliers, namely the entries labelled in Ref.~\cite{pdg2020} as ACHASOV18, AKHMETSHIN00D, ANTONELLI88, FUKUI88 
(i.e., four out of the five datapoints obtained from the $\eta \rho^0$ mode, all corresponding to surprisingly low estimates), BARTOS17, LEES17C, BERTIN98 \& 97D (from the $\pi \pi$ mode), and ABELE99D (from the $K \bar{K}$ 
mode).
\item The trimmed dataset was submitted to the optimisation using the standard $\chi^2$ MF and resulted in $\chi^2_{\rm min} \approx 72.01$ for $23$ DoF, i.e., in a Birge factor of about $1.77$.
\item The results from the four statistical methods of this work, as well as the PDG value, agree within the uncertainties, thus suggesting that the total decay width of the $\rho(1450)$ is large, in the vicinity of $400$ 
MeV. The relevant uncertainty in the recommended value of this work (result with Andrews weights) is about $30~\%$ smaller than the one in the PDG estimate.
\end{itemize}

The decay mode of the $\rho(1450)$ resonance to two pions is marked in Ref.~\cite{pdg2020} as `seen', but the sparseness of the relevant data does not enable a serious determination.

\subsubsection{\label{sec:rho1570}$\rho(1570)$}

Little is known about this resonance, in practice stemming from a 2008 report by the BaBar Collaboration. The PDG recommend the use of $1570(36)(62)$ MeV for the mass and $144(75)(43)$ MeV for the total decay width. Regarding 
the decay modes of the $\rho(1570)$, two have been experimentally established (to the $e^+ e^-$ and $\omega \pi$ final states). The two-pion decay of the $\rho(1570)$ has not been documented.

\subsubsection{\label{sec:rho1700}$\rho(1700)$}

\begin{figure}
\begin{center}
\includegraphics [width=15.5cm] {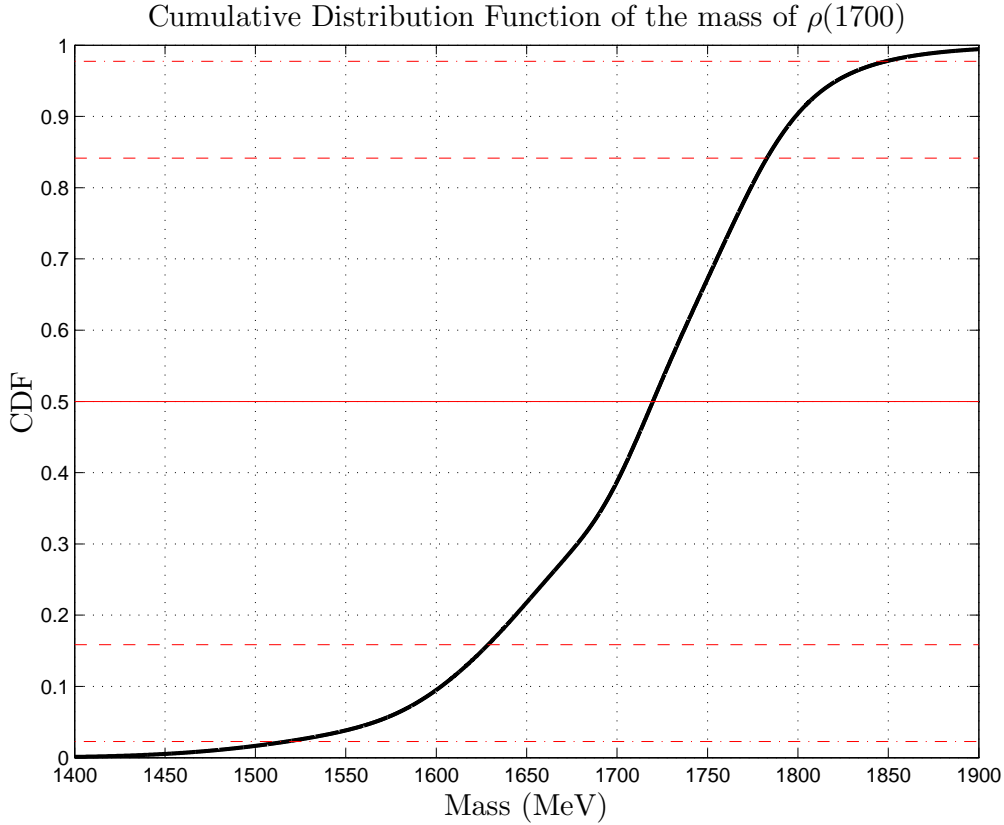}
\caption{\label{fig:MassOfrho1700}The CDF of the mass of the $\rho(1700)$, obtained from averaging fifteen Gaussian distributions. The horizontal straight lines represent the same levels as in Fig.~\ref{fig:MassOfrho770}.}
\vspace{0.35cm}
\end{center}
\end{figure}

As the PDG base their estimates for the mass and for the total decay width of the $\rho(1700)$ on the data from the $\eta \rho^0$ and $\pi^+ \pi^-$ modes, only these two datasets will be imported in this work. Regarding the 
mass, the results are shown in Table \ref{tab:rho1700}, upper part.
\begin{itemize}
\item The application of the two robust methods to the original input dataset of $23$ datapoints yielded eight outliers, labelled in Ref.~\cite{pdg2020} as ACHASOV18 (from the $\eta \rho^0$ mode), and BARTOS17, LEES12G, 
GESHKENBEIN89, ASTON80, ATIYA79B, BECKER79, and HYAMS73 (from the $\pi^+ \pi^-$ mode).
\item The trimmed dataset was submitted to the optimisation using the standard $\chi^2$ MF and resulted in $\chi^2_{\rm min} \approx 47.17$ for $14$ DoF, i.e., in a Birge factor of about $1.84$.
\item The results from the four statistical methods of this work agree among themselves, as well as with the PDG estimate.
\item The CDF of the mass of the $\rho(1700)$ is displayed in Fig.~\ref{fig:MassOfrho1700}.
\end{itemize}

\begin{table}%[h!]
{\bf \caption{\label{tab:rho1700}}}The equivalent of Table \ref{tab:rho770} for the $\rho(1700)$. Regarding the methods featuring the $\chi^2$ MF and the CDF, eight outliers in case of the mass and three in case of the total 
decay width, identified as such by the robust-optimisation methods of Andrews and Tukey, were removed from the original input dataset.
\vspace{0.3cm}
\begin{center}
\begin{tabular}{|c|c|}
\hline
Source/Method & Result\\
\hline
\hline
\multicolumn{2}{|c|}{Mass (MeV)}\\
\hline
PDG \cite{pdg2020} & $1720 \pm 20$\\
Andrews weights & $1723^{+29}_{-28}$\\
Tukey weights & $1723^{+29}_{-28}$\\
Standard $\chi^2$ & $1717 \pm 12$\\
CDF & $1720^{+16}_{-24}$\\
\hline
\multicolumn{2}{|c|}{Total decay width (MeV)}\\
\hline
PDG \cite{pdg2020} & $250 \pm 100$\\
Andrews weights & $271 \pm 15$\\
Tukey weights & $270 \pm 15$\\
Standard $\chi^2$ & $261 \pm 10$\\
CDF & $263^{+17}_{-23}$\\
\hline
\multicolumn{2}{|c|}{Branching fraction of $\rho(1700) \to \pi \pi$ ($\%$)}\\
\hline
PDG \cite{pdg2020} & $-$\\
Andrews weights & $27.4^{+2.6}_{-2.7}$\\
Tukey weights & $27.4^{+2.6}_{-2.7}$\\
Standard $\chi^2$ & $27.4 \pm 2.5$\\
CDF & $27.0^{+2.8}_{-3.2}$\\
\hline
\end{tabular}
\end{center}
\vspace{0.5cm}
\end{table}

\begin{figure}
\begin{center}
\includegraphics [width=15.5cm] {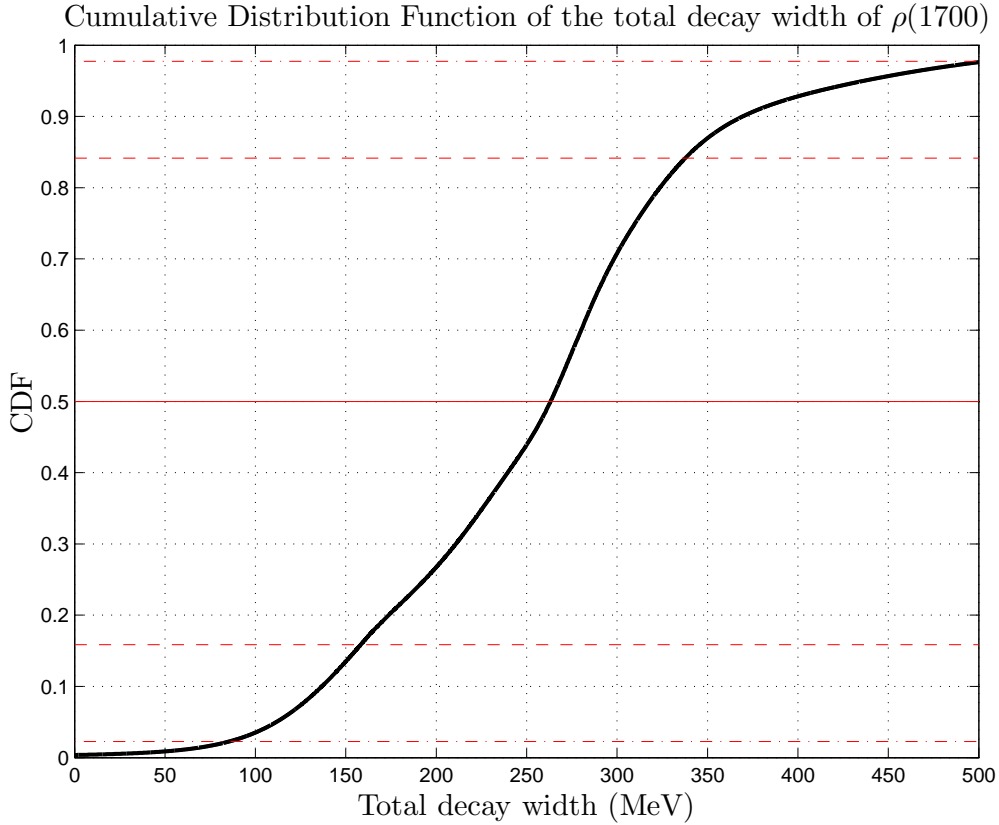}
\caption{\label{fig:WidthOfrho1700}The CDF of the total decay width of the $\rho(1700)$, obtained from averaging twenty Gaussian distributions. The horizontal straight lines represent the same levels as in Fig.~\ref{fig:MassOfrho770}.}
\vspace{0.35cm}
\end{center}
\end{figure}

For the total decay width of this resonance, the results are shown in Table \ref{tab:rho1700}, middle part.
\begin{itemize}
\item The application of the two robust methods to the original input dataset of $23$ datapoints yielded three outliers, namely the entries labelled in Ref.~\cite{pdg2020} as BARTOS17A ($489.58 \pm 16.95$ MeV), LEES17C, and 
GESHKENBEIN89 (all from the $\pi \pi$ mode).
\item The trimmed dataset was submitted to the optimisation using the standard $\chi^2$ MF and resulted in $\chi^2_{\rm min} \approx 46.51$ for $19$ DoF, i.e., in a Birge factor of about $1.56$.
\item The results from the four statistical methods of this work agree among themselves, as well as with the PDG estimate, which is accompanied by a significantly larger uncertainty.
\item The CDF of the total decay width of the $\rho(1700)$ is displayed in Fig.~\ref{fig:WidthOfrho1700}.
\item The available data on the total decay width of this resonance provide an explanation of the general difficulty in extracting reliable estimates for the physical properties of the scalar-isoscalar and vector-isovector 
mesons. The original input dataset for the total decay width of the $\rho(1700)$ contains largely incompatible values, even when they originate from reports by the same first author, published in the same year; for the sake 
of example, the entries labelled as BARTOS17 ($268.98 \pm 11.40$ MeV) and BARTOS17A ($489.58 \pm 16.95$ MeV) are nearly $11 \sigma$ apart (combined uncertainty). To accommodate in the same fit datapoints as the aforementioned 
two appears to be an onerous, if not an impossible, task, at least when intending to make exclusive use of conventional statistical methods.
\end{itemize}

Four estimates are available for the branching fraction of the $\rho(1700)$ decay to two pions. The datapoint labelled in Ref.~\cite{pdg2020} as MARTIN78C (described therein as representing the interval `from 0.15 to 0.30') 
is taken to correspond to the value $0.225 \pm 0.075$. The application of the four statistical methods of this work yields consistent results (see Table \ref{tab:rho1700}, lower part) and no outliers. Although the PDG made 
no attempt to extract a result from these four values, an estimate of about $27~\%$, with an uncertainty of about $3~\%$, may be obtained from the available data.

\subsubsection{\label{sec:rho1900}$\rho(1900)$}

Few results for the mass and for the total decay width of this resonance are available, five in each case. The PDG do not attempt to extract estimates from these data.

For the mass of this resonance, the results of this work are given in Table \ref{tab:rho1900}, upper part.
\begin{itemize}
\item The application of the two robust methods to the original input dataset yielded no outliers.
\item The original dataset was submitted to the optimisation using the standard $\chi^2$ MF and resulted in $\chi^2_{\rm min} \approx 10.58$ for $4$ DoF, i.e., in a Birge factor of about $1.63$ and p-value of about 
$3.17 \cdot 10^{-2}$, i.e., in an acceptable outcome in terms of the significance level of this work.
\end{itemize}

\begin{table}%[h!]
{\bf \caption{\label{tab:rho1900}}}The equivalent of Table \ref{tab:rho770} for the $\rho(1900)$.
\vspace{0.3cm}
\begin{center}
\begin{tabular}{|c|c|}
\hline
Source/Method & Result\\
\hline
\hline
\multicolumn{2}{|c|}{Mass (MeV)}\\
\hline
PDG \cite{pdg2020} & $-$\\
Andrews weights & $1880 \pm 22$\\
Tukey weights & $1879^{+23}_{-21}$\\
Standard $\chi^2$ & $1887 \pm 10$\\
CDF & $1883^{+15}_{-12}$\\
\hline
\multicolumn{2}{|c|}{Total decay width (MeV)}\\
\hline
PDG \cite{pdg2020} & $-$\\
Andrews weights & $151^{+73}_{-75}$\\
Tukey weights & $151^{+73}_{-75}$\\
\hline
\end{tabular}
\end{center}
\vspace{0.5cm}
\end{table}

For the total decay width of this resonance, the results are given in Table \ref{tab:rho1900}, lower part. The application of the two robust methods to the original input dataset yielded three outliers, namely the entries 
labelled in Ref.~\cite{pdg2020} as AUBERT08S, FRABETTI04, and ANTONELLI96 (all of which correspond to average values below $100$ MeV). Given that the optimisation suggests the removal of more than half of the input data, its 
outcome is deemed unreliable. Notwithstanding, it found its way into Table \ref{tab:rho1900} on account of the reasoning that the uncertainties in the extracted estimates are so large that they practically encompass all 
values or, better said, they are not incompatible with any.

Finally, the two-pion decay of the $\rho(1900)$ has not been experimentally established. The sparseness of the information on the physical properties of this resonance, as well as the lack of the experimental evidence on its 
decay into two pions, precludes any plans to include the corresponding $t$-channel exchange graph in the $\pi N$ interaction model of Ref.~\cite{matsinos2017}.

\subsubsection{\label{sec:ConclusionsVectorIsovector}Concluding remarks on the vector-isovector mesons}

The values of the $W$-statistic and the p-values from the Shapiro-Wilk test in case of the masses and total decay widths of the five vector-isovector mesons of this section can be found in Table \ref{tab:SWTestVectorIsovector}. 
In four cases (i.e., for the mass of the $\rho^0(770)$, and for the total decay widths of the $\rho^\pm(770)$, of the $\rho^0(770)$, and of the $\rho(1450)$), the p-values drop below $10~\%$, suggesting sizeable, yet not 
unacceptable (at the significance level of this work), deviations from the normal distribution.

\begin{table}%[h!]
{\bf \caption{\label{tab:SWTestVectorIsovector}}}The results of the application of the Shapiro-Wilk test \cite{Shapiro1965,Royston1995} to the normalised residuals of the $\chi^2$ fits to the trimmed datasets containing 
the estimates for the mass and for the total decay width of each vector-isovector meson. The $\rho^\pm(770)$ and the $\rho^0(770)$ are separately treated. In case of the $\rho(1570)$, the test cannot be performed (as only 
one input result is available). In case of the total decay width of the $\rho(1900)$, the $\chi^2$ fit was not attempted, as the robust methods suggested the removal of three, out of the five available, datapoints of the 
original input dataset.
\vspace{0.3cm}
\begin{center}
\begin{tabular}{|c|c|c|}
\hline
Case & $W$-statistic & p-value\\
\hline
\hline
Mass of the $\rho^\pm(770)$ & $0.960$ & $0.798$\\
Total decay width of the $\rho^\pm(770)$ & $0.810$ & $2.68 \cdot 10^{-2}$\\
\hline
Mass of the $\rho^0(770)$ & $0.862$ & $5.24 \cdot 10^{-2}$\\
Total decay width of the $\rho^0(770)$ & $0.898$ & $8.91 \cdot 10^{-2}$\\
\hline
Mass of the $\rho(1450)$ & $0.972$ & $0.686$\\
Total decay width of the $\rho(1450)$ & $0.902$ & $2.33 \cdot 10^{-2}$\\
\hline
Mass of the $\rho(1700)$ & $0.939$ & $0.375$\\
Total decay width of the $\rho(1700)$ & $0.937$ & $0.211$\\
\hline
Mass of the $\rho(1900)$ & $0.947$ & $0.715$\\
Total decay width of the $\rho(1900)$ & $-$ & $-$\\
\hline
\end{tabular}
\end{center}
\vspace{0.5cm}
\end{table}

\section{\label{sec:Conclusions}Conclusions}

The pion-nucleon ($\pi N$) interaction model of Ref.~\cite{matsinos2017} is based on $s$-, $u$-, and $t$-channel exchanges of hadrons. Regarding its $t$-channel Feynman graphs, relevant are the scalar-isoscalar 
$I^G\,(J^{PC}) = 0^+\,(0^{++})$ and the vector-isovector $I^G\,(J^{PC}) = 1^+\,(1^{--})$ mesons with rest masses below $2$ GeV; there are five such resonances in each category. The extraction of the important (in the context 
of this model) physical properties of these resonances (i.e., estimates for the masses and for the partial decay widths to two pions) is the objective of this study.

The main results are obtained via the application of two robust-optimisation methods, variants of the standard $\chi^2$ method, and differing from it on account of a continuous weight, which is applied to each input datapoint 
on the basis of its distance to the bulk of the data. Selected in this work are two types of weights (Andrews and Tukey) which provide maximal insensitivity to the presence of discrepant observations (outliers) among the 
input data (by admitting constant contributions from these datapoints to the MF). The input data comprise the entirety of the datasets (results accompanied by uncertainties) listed in the recent compilation by the Particle 
Data Group (PDG) \cite{pdg2020}; regarding their occasional selection of specific modes as better-suited for the extraction of reliable estimates, their recommendations have been followed.

In comparison with conventional statistical methods, robust techniques offer distinct advantages in tackling input datasets with outliers. To start with, such datapoints need not be removed from the input dataset during the 
optimisation: the application of the distance-dependent weights of Eqs.~(\ref{eq:EQ09_01}-\ref{eq:EQ09_07}) reduces their contributions to the MF, and renders them harmless. Furthermore, the data themselves decide which input 
datapoint is a regular observation and which an outlier at any given step of the optimisation.

Results were also obtained after using two non-robust processing methods, i.e., the optimisation based on the standard $\chi^2$ MF and the analysis featuring the determination of the physical properties of the resonances 
from the CDF, a method which had been applied to discrepant data in Ref.~\cite{Matsinos2019}. To have confidence in the results, any outliers (identified as such by the two robust methods) were removed from each original 
input dataset when using these two methods. The estimates obtained via the application of the two non-robust methods are considered supplementary, serving the purpose of testing the self-consistency of this study. The 
results, obtained from the four statistical methods of this work, were found compatible among themselves.

The first part of the analysis (Section \ref{sec:ScalarIsoscalar}) relates to the scalar-isoscalar mesons; a summary of the results is given in Table \ref{tab:ScalarIsoscalar}. Regarding the masses and the total decay widths 
of these resonances, $437$ values are listed in Ref.~\cite{pdg2020}, yet only $28$ are properly analysed therein (i.e., resulting in the determination of an average and of a relevant meaningful uncertainty in each case). Most 
of the available data were treated by the PDG as sources of rough guesses as to the significant domain of the various PDFs, yielding only broad recommended ranges of probable values, as the case is for the masses and for the 
total decay widths of the $f_0(500)$, of the $f_0(980)$, and of the $f_0(1370)$. The PDG provide proper results for the $f_0(1500)$ and for the $f_0(1710)$ on the basis of $28$ out of the $201$ datapoints for these two 
resonances. Regarding the branching fractions to two pions (and excluding the evident result for the $f_0(500)$), only in case of the $f_0(1500)$ is a proper estimate given in Ref.~\cite{pdg2020}, though that estimate was 
based on ten (out of the $25$ available) datapoints, two of which were established as outliers herein.

On the contrary, all available data were analysed in this work using the two robust methods; I will mention once again that no data need be discarded when applying robust techniques, hence the main results of this work 
contain the effects of the outliers (albeit in a reduced form). Before submitting the data to the processing with the non-robust methods featuring the $\chi^2$ MF and the CDF, $69$ datapoints were removed; this corresponds 
to a data-rejection ratio of about $15.8~\%$. This work provides proper results for the masses and for the total decay widths of all five scalar-isoscalar resonances. Regarding the branching fractions to two pions, only in 
case of the $f_0(1370)$ is no result mentioned, as (given the poor status of the available information regarding that physical quantity) there can be no confidence in the extracted estimate (see concluding remarks in Section 
\ref{sec:f01370}). To summarise, this work demonstrates that the proper analysis of all available data can lead to the determination of fourteen (out of fifteen) quantities needed in the context of the $\pi N$ interaction 
model of Ref.~\cite{matsinos2017}.

The second part of the analysis (Section \ref{sec:VectorIsovector}) deals with the vector-isovector mesons; a summary of the results is given in Table \ref{tab:VectorIsovector}. A distinction is made among the members of the 
$\rho(770)$ isospin triplet, providing estimates both for the $\rho^\pm(770)$ as well as for the $\rho^0(770)$. Regarding the masses and the total decay widths of the vector-isovector resonances, $183$ values are listed in 
Ref.~\cite{pdg2020}, yet only $24$ are properly analysed therein. The PDG provide proper results only for the $\rho(770)$. They also accept one of the two available estimates in case of the mass of the $\rho(1570)$; 
similarly, for the total decay width of that resonance.

Again, all available data were analysed in this work using the two robust methods. Before submitting the data to the processing with the non-robust methods, $45$ datapoints were removed; this corresponds to a data-rejection 
ratio of about $24.6~\%$. This work provides proper results for the masses and for the total decay widths of all vector-isovector resonances, save for the $\rho(1570)$ where the only two available results for the mass (and 
another two for the total decay width) preclude a statistical analysis. Regarding the branching fractions of the decay to two pions, estimates are obtained from the available data in the cases of the $\rho(770)$ and of the 
$\rho(1700)$. In the former case, the PDG recommend the use of $\approx 100~\%$; in the latter, they did not extract an estimate. It must be borne in mind that the two-pion decay mode has not been established in two cases, 
for the $\rho(1570)$ and for the $\rho(1900)$, whereas it has been observed in case of the $\rho(1450)$, yet no estimates for the relevant branching fraction appear listed in Ref.~\cite{pdg2020}.

This report leaves no doubt that the uncertainties, accompanying the various determinations of the masses and of the total decay widths of the scalar-isoscalar and vector-isovector mesons below $2$ GeV (as they appear in 
Ref.~\cite{pdg2020}), have been underestimated, by no less than $40~\%$ \emph{on average} (see also the discussion in Section \ref{sec:MethodsI}). Only in two cases, namely in those relating to the masses of the $\rho(770)$, 
are Birge factors $s<1$ obtained, yet the surprising smallness of the corresponding $\chi^2_{\rm min}$ results leaves unanswered questions about the uncertainties of the $21$ input datapoints. The large amount of identified 
outliers among the $437+183=620$ input data also points in the direction of underestimated uncertainties. Assuming pure statistical fluctuations, the robust-optimisation method of Andrews should have identified 
$2.59 \cdot 10^{-5} \times 620 \approx 0.016$ outliers; identified instead were $114$.

I will finally express my opinion clearly, for what it is worth. Disregarding all available data on a physical quantity and opting for a rough determination of the `region of interest' regarding that quantity is a peculiar 
practice, in particular when it is not accompanied by sufficient explanation of why this is inevitable. Furthermore, such a practice is discouraging, and it does not do justice to the experimenters. If the lack of details on 
the rationale behind such a decision should be interpreted as criticism of the disagreement among the available results, then such an unexpressed, indirect criticism is unconstructive as it does nothing to motivate additional 
experimental activity, aiming at resolving any potential conflicts.

\begin{ack}
The figures of this paper were created with MATLAB$^{\textregistered}$ (The MathWorks, Inc., Natick, Massachusetts, United States).
\end{ack}

\end{document}